\documentclass[11pt]{article}

\usepackage{thmtools}
\usepackage{thm-restate}
\usepackage{cleveref}

\usepackage{mathrsfs}           
\usepackage{vmargin,fancyhdr}   
\usepackage{algorithm}
\usepackage{algorithmic}

\usepackage{epstopdf}
\usepackage{graphicx}

\usepackage{enumerate}

\usepackage{amsmath,amssymb}    
\usepackage{verbatim}           
\usepackage{xspace}             
\usepackage{graphicx,float}     
\usepackage{ifthen,calc}        
\usepackage{textcomp}           
\usepackage{fancybox}           
\usepackage{hhline}             
\usepackage{float}              

\usepackage[rflt]{floatflt}     
\usepackage[small,compact]{titlesec}
\usepackage{setspace}
\usepackage{subfigure}

\newcommand{\proof}[0]{\smallskip\noindent\textit{\textbf{Proof}}\quad}
\newcommand{\Proof}[0]{\smallskip\noindent\textit{\textbf{Proof}}\quad}

\newtheorem{theorem}{Theorem}[section]

\newtheorem{corollary}[theorem]{Corollary}
\newtheorem{definition}[theorem]{Definition}

\newtheorem{remark}[theorem]{Remark}
\newtheorem{lemma}[theorem]{Lemma}

\newcommand{\QED}[0]{\hfill\ensuremath{\blacksquare}\medspace}

\usepackage{times}

\begin{document}

\title {Stochastic Block Model and Community Detection in the Sparse Graphs: A spectral algorithm with optimal rate of recovery}


\author{
Peter Chin\\
Boston University and Draper laboratory \thanks{This work was supported by NSF-DMS-1222567 from NSF and FA9550-12-1-0136 from the Air Force Office of Scientific Research. A. Rao is supported by NSF grant CCF-1111257.} \\
\texttt{spchin@cs.bu.edu}
\and
  Anup Rao\\
  Yale University \thanks{This work was partially supported by NSF grant CCF-1111257.} \\
  \texttt{anup.rao@yale.edu}
 \and 
   Van Vu\\
  Yale University \thanks{This work was supported by research grants DMS-0901216 and AFOSAR-FA-9550-09-1-0167 and a grant from Draper Lab.}\\
  \texttt{van.vu@yale.edu}
}

\maketitle

%

\maketitle

\begin{abstract}
In this paper, we present  and analyze  a simple  and robust  spectral algorithm 
for the stochastic block model with $k$ blocks, for any $k$ fixed.  Our algorithm works with graphs having constant edge density, 
under an optimal condition on the gap between the density inside a block and the density between the blocks. 
As a co-product, we settle an open question posed by Abbe et. al. concerning  censor block models. 
\end{abstract}

\def\bvec#1{{\mbox{\boldmath $#1$}}}

\renewcommand\aa{\boldsymbol{\mathit{a}}}
\newcommand\bb{\boldsymbol{\mathit{b}}}
\newcommand\vv{\boldsymbol{\mathit{v}}}
\newcommand\uu{\boldsymbol{\mathit{u}}}
\newcommand\dd{\boldsymbol{\mathit{d}}}
\newcommand\ee{\boldsymbol{\mathit{e}}}
\newcommand\ff{\boldsymbol{\mathit{f}}}
\newcommand\rr{\boldsymbol{\mathit{r}}}
\newcommand\error{\boldsymbol{\mathit{err}}}
\newcommand\xx{\boldsymbol{\mathit{x}}}
\newcommand\yy{\boldsymbol{\mathit{y}}}
\newcommand\zz{\boldsymbol{\mathit{z}}}
\newcommand\xxbar{\overline{\boldsymbol{\mathit{x}}}}

\newcommand\ww{\boldsymbol{\mathit{w}}}

\newcommand{\erExact}{\ensuremath{\overline{\boldsymbol{\tau}}}}
\newcommand{\er}{\ensuremath{\boldsymbol{\tau}}}

\renewcommand\AA{{\mathit{A}}}
\newcommand\CC{{\mathit{C}}}
\newcommand\BB{{\mathit{B}}}
\newcommand\HH{{\mathit{H}}}
\newcommand\II{{\mathit{I}}}
\newcommand\LL{{\mathit{L}}}
\newcommand\MM{{\mathit{M}}}
\newcommand\RR{{\mathit{R}}}
\newcommand\WW{{\mathit{W}}}
\newcommand\XX{{\mathit{X}}}
\newcommand\YY{{\mathit{Y}}}
\newcommand\ZZ{{\mathit{Z}}}

\newcommand\E{{\boldsymbol{E}}}
\newcommand\var{{\boldsymbol{\mathit{Var}}}}

\newcommand\PPi{{\Pi}}

\def\norm#1{\left\| #1 \right\|}

\newcommand\eFlow[1]{\mathcal{E}_{\RR}(#1)}

\newcommand{\defeq}{\buildrel \text{d{}ef}\over =}
\newcommand{\Oh}{\ensuremath{\mathcal{O}}}
\newcommand{\oh}{\ensuremath{\mathcal{O}}}
\newcommand{\hsum}[2]{\operatorname{HrmSum}{\ensuremath{\left(#1, #2\right)}}}

\def\prob#1#2{\mbox{Pr}_{#1}\left[ #2 \right]}
\newcommand{\nbr}[1]{\left\|#1\right\|}
\newcommand{\trace}[1]{\operatorname{Tr}\left( #1 \right)}
\newcommand{\func}[1]{f_{#1}}
\newcommand{\cexpct}[3]{\ensuremath{\mathbb{E}}_{#1}\left[#2\vert#3\right]}

\newcommand{\poly}{\ensuremath{\textbf{poly}}}

\newcommand{\mati}{\ensuremath{I}}
\newcommand{\laplacian}{\ensuremath{L}}
\newcommand{\mate}{\ensuremath{E}}
\newcommand{\mata}{\ensuremath{A}}

\newcommand{\expct}[1]{\ensuremath{\mathbb{E}}{#1}}
\newcommand{\matd}{\ensuremath{D}}
\newcommand{\matb}{\ensuremath{B}}
\newcommand{\matx}{\ensuremath{X}}
\newcommand{\maty}{\ensuremath{Y}}
\newcommand{\matz}{\ensuremath{Z}}
\newcommand{\matu}{\ensuremath{U}}
\newcommand{\mathh}{\ensuremath{H}}
\newcommand{\matnx}{\ensuremath{S}}
\newcommand{\matny}{\ensuremath{R}}
\newcommand{\matni}{}

\newcommand{\vecindicator}{\ensuremath{\boldsymbol{\chi}}}
\newcommand{\vecb}{\ensuremath{\textbf{b}}}
\newcommand{\err}{\ensuremath{\textbf{err}}}
\newcommand{\vecr}{\ensuremath{\textbf{r}}}
\newcommand{\vecu}{\ensuremath{\textbf{u}}}
\newcommand{\vecnnv}{\ensuremath{\textbf{z}}}
\newcommand{\vecv}{\ensuremath{\textbf{v}}}
\newcommand{\vecx}{\ensuremath{\textbf{x}}}
\newcommand{\vecy}{\ensuremath{\textbf{y}}}

\section{Introduction}

Community detection is an important problem  in statistics, theoretical computer science and image processing. A widely studied theoretical model in this area is the stochastic block model.  In the simplest case, there are two blocks $V_1, V_2$ each of size of $n$;  one considers a random graph generated from the following distribution: an edge between vertices belonging to the same block appears with probability $\frac{a}{n}$ and an edge between vertices across different blocks appear with probability $\frac{b}{n}$, where $a > b >0$. Given an instance of this graph, we would like to identify the two blocks as correctly as possible. 
Our paper will deal with the general case of $k \ge 2 $ blocks, but for the sake of simplicity, let us first focus on  $k=2$.


For $k=2$, the  problem can be seen as a variant of  the well known hidden bipartition problem, which has been studied 
by many researchers   in theoretical computer science, starting with the work of \cite{Bui}; 
(see \cite{Dyer} \cite{RaviFOCS28} \cite{Jerrum} \cite{McSherry} and the references therein 
for further developments).  In these earlier papers, $a$ and $b$ are large (at least $\log n$) and the goal is to recover both blocks completely.
It is known  that one can efficiently obtain a complete recovery  if $\frac{(a-b)^2}{a+b}  \ge C \frac{\log n}{n} $ and $a, b \ge C \log n$ for some sufficiently large constant $C$ (see, for instance \cite{2014Vu}). 

  In the stochastic block model problem, the graph is sparse with $a$ and $b$ being  constants. Classical results from random graph theory tell us  that in this range the graph contains, with high probability,  a linear portion of isolated vertices \cite{bollobas}. Apparently, there is no way to tell these vertices apart  and so a complete recovery is out of  question. 
The  goal here is  to recover a large portion  of each block, namely finding a partition $V_1 ' \cup V_2' $ of $V= V_1 \cup V_2$ such that 
$V_i$ and $V_i'$ are close to each other.   For quantitative purposes, let us introduce a definition

\begin{definition}  A  collection of subsets $V_1', V_2 ' $ of $V_1 \cup V_2$  is $\gamma$-correct if $|V_i \cap V_i'| \ge (1- \gamma) n $.  \end{definition}

\noindent  In \cite{CPC:7183040}, Coja-Oglan proved 
  
  \begin{theorem} \label{OJ}  For any constant $\gamma >0$ there are constants $d_0 , C >0$ such that  if 
   $a,b >d_0$ and $\frac{(a-b)^2 }{a+b} >C\log (a+b) $, one can find a $\gamma $-correct partition using a  polynomial time algorithm. 
   \end{theorem}

  Coja-Oglan  proved Theorem \ref{OJ}  as part of a more general problem, and his  algorithm was rather involved. 
  Furthermore, the result is not yet sharp and it  has been conjectured that the  $\log $ term is removable\footnote{We would like to thank E. Abbe for communicating this conjecture.}. 
  Even when the log term is removed, an important question is to find out the optimal  relation between the accuracy $\gamma$ and the ratio $\frac{(a-b)^2 }{a+b} $. This is the main goal of this paper.


\begin{theorem}
\label{thm:main1} There are constants $C_0$ and $C_1$ such that the following holds. For any constants $a> b >C_0$ and $\gamma > 0$ satisfying  

$$\frac{(a-b)^2 }{ a+b}  \geq C_1 \log \frac{1}{\gamma}, $$
 we can find a $\gamma $-correct partition with probability $1-o(1)$  using a simple spectral algorithm. 
 \end{theorem}

 The  constants $C_0,C_1$ can be computed explicitly via a careful, but rather tedious, book keeping. We try not to optimize these constants 
to simplify the presentation.  The proof of Theorem \ref{thm:main1} yields the following corollary 

\begin{corollary}
\label{cor:main1} There are constants $C_0$ and $\epsilon$ such that the following holds. For any constants $a> b >C_0$ and $\epsilon > \gamma > 0$ satisfying  

$$\frac{(a-b)^2 }{ a+b}  \geq  8.1 \log \frac{2}{\gamma}, $$
 we can find a $\gamma $-correct partition with probability $1-o(1)$  using a simple spectral algorithm. 
 \end{corollary}

 In parallel to our study, \cite{Zhou} , proving a minimax rate result that suggested that there is a constant $c>0$  
  $$\frac{(a-b)^2} {a+b}  \le c  \log \frac{1}{\gamma}   $$

\noindent  then one {\it cannot }  recover  a $\gamma$-correct partition (in expectation), regardless the algorithm.

 In order to prove  Theorem \ref{thm:main1}, we  design a fast and robust  algorithm which obtains a $\gamma$-correct partition under the 
  condition $\frac{(a-b)^2} {a+b}  \ge C \log \frac{1}{\gamma} $. Our algorithm guarantees  $\gamma$-correctness with high probability.

 We can refine the algorithm to handle the (more difficult) general case of having  $k$ blocks, for any fixed number $k$. Suppose now there are $k$ blocks 
  $V_1,...,V_k$ with $|V_i| =\frac{n}{k}$ with edge probabilities $\frac{a}{n}$ between vertices within the same block and $\frac{b}{n}$ between vertices in different blocks. As before, a  collection of subsets $V_1', V_2 ',..,V_k' $ of $V_1 \cup V_2 \cup ... \cup V_k$  is $\gamma$-correct if $|V_i \cap V_i'| \ge (1- \gamma) \frac{n}{k} $.

\begin{theorem} 
\label{them:multComm}
 There exists constants $C_1,C_2$, such that if  $k$ is any constant as $n \to \infty$
  and if 
\begin{enumerate}
\item $a > b \geq  C_1$ 
\item $(a-b)^2 \geq C_2 k^2 a \log \frac{1}{\gamma} $,
\end{enumerate} then we can find a $\gamma$-correct partition with probability at least $1-o(1)$ using a simple spectral algorithm. 
\end{theorem} 

We believe that this result is sharp, up to the values of $C_1$ and $C_2$; in particular, the requirement $\frac{(a-b)^2}{a } = \Omega (k^2)$ is optimal.

Our method also works (without significant changes) in the case the blocks
are not equal, but have comparable sizes (say $cn \ge |V_i | \ge n$ for some constant $c \ge 1$). In this case, the constants $C_1, C_2$ above will also depend on $c$.  While the emphasis of this paper is on the case $a,b$ are constants, 
we would like to point out that this assumption is not required in our theorems, so our algorithms work on denser graphs as well. 

Let us now  discuss some recent works, which we just learned after posting the first version of this paper on arxiv. 
 Mossel  informed us about a  recent  result  in \cite{MosselOpt}  which is similar to  Theorem \ref{thm:main1} (see \cite[Theorem 5.3]{MosselOpt}). They give a polynomial time algorithm and prove that there exists a constant $C$ such that if $(a-b)^2 > C (a+b)$ and $a,b$ are fixed as $n \to \infty$, then the algorithm recovers an optimal fraction of the vertices.  The algorithm in \cite{MosselOpt}  is very different from ours, and uses non-back tracking walks. This algorithm doest not yet handle the case 
  of more than 2 blocks, and its  analysis looks very delicate. Next,  Guedon sent us \cite{GV}, in which the authors  also proved a result similar to Theorem \ref{thm:main1},  under a stronger  assumption
  $ (a-b)^2 \ge C \frac{1}{\gamma^2}  (a+b) $ (see Theorem 1.1 and Corollary 1.2 of \cite{GV}). Their approach relies on an entirely different (semi-definite program)  algorithm, which, in turn,  was   based on Grothendick's inequality. 
  This approach seems to extend to the general $k >2$ case; however, the formulation of the result in this case, using matrix approximation,  is somewhat different from ours (see \cite[Theorem 1.3]{GV}). 
  Two more closely related papers have been  brought to our attention by the reviewers. In \cite{Mas13}, authors have worked out the spectral part of the result in this paper. 

 It is  remarkable to see so many progresses, using different approaches,  on the same problem in such a short  span of time.   This suggests  that the problem is  indeed 
 important and rich, and it  will be really pedagogical to study the performance of  the existing  algorithms in practice. We are going to discuss the performance of our algorithm in Sections 3 and 4. 

We next  present  an application of our method to the {\it Censor Block Model} studied by  Abbe et. al. in \cite{Abbe}. 
As before, let $V$ be the union of two blocks $V_1, V_2$, each of size $n$. 
Let $G=(V,E)$ be a random graph with edge probability $p$  with incidence  matrix $B_G$ and $\xx = (x_1,...,x_{2n})$ be the indicator vector of  $V_2$. 
Let $\zz$ be a random noise vector whose coordinates  $z_{e_i}$ are i.i.d $\text{Bernoulli}(\epsilon)$ (taking value $1$ with probability $\epsilon$ and $0$ otherwise), where $e_i$ are the edges of 
$G$.

Given a noisy observation $$\yy = B_G \xx \oplus \zz $$ where $\oplus$ is the addition in mod $2$, one would like identify the blocks.  In \cite{Abbe}, the authors proved that exact recovery ($\gamma =0$) is possible if and only if $\frac{np}{\log n} \geq \frac{2}{(1-2\epsilon)^2} + o(\frac{1}{(1-2\epsilon)^2})$ in the limit $\epsilon \to 1/2$. Further, they gave a semidefinite programming based algorithm which succeeds up to twice the threshold. They posed the question of partial recovery ($\gamma >0$)   for sparse graphs.   
  Addressing this question, we show
  
  \begin{theorem}
  \label{thm:CBM}
  For any given constants $\gamma , 1/2 > \epsilon > 0$, there exists  constant $C_1,C_2$ such that if  $np \geq \frac{C_1}{(1-2\epsilon)^2}$ and $p \geq \frac{C_2}{n}$, then 
   we can find a $\gamma$-correct partition with probability $1-o(1)$, using a simple spectral algorithm. 
  \end{theorem}

Let us conclude this section by mentioning  a related, interesting, problem, where the purpose is just to do better than a random guess (in our terminology, to find a partition 
which is $(1/2 +\epsilon ) $-correct). It was  conjectured in~\cite{PhysRevE.84.066106} that this is possible if and only if  $(a-b)^2 > (a+b)$. This conjecture has been settled recently by  Mossel et. al. \cite{2012arXiv1202.1499M} \cite{2013arXiv1311.4115M} and Massoulie 
\cite{Massoulie13} .  Another closely related problem which has been studied in  \cite{AbbeComp} \cite{MNSComp} is about when one can recover at least $1-o(1)$ fraction of the vertices.

\vskip2mm 
 
 The rest of the paper is organized as follows.  In section~\ref{sec:Algo}, we  describe our algorithm for Theorem~\ref{thm:main}  and  an overview of the proof.
 The full proof comes in sections \ref{sec:Proof}.  In section \ref{sec:MultComm}, we show how to modify the algorithm to handle the $k$ block case and prove theorem \ref{them:multComm}. Finally, in section \ref{sec:CBM}, we prove theorem \ref{thm:CBM}.

\section{Two communitites} 

We first consider the case $k=2$. Our algorithm will have two steps. First we use  a spectral algorithm to recover a partition  where the dependence between $\gamma$ and $\frac{(a-b)^2 }{a+b} $ is sub-optimal. 

\label{sec:Algo}

 Let $\mata_0$ denote the adjacency matrix of a random graph generated from the distribution as in Theorem \ref{thm:main}. Let $\bar{\mata}_0 \defeq \expct{\mata_0}$ and $\mate_0 \defeq \mata_0 - \bar{\mata}_0$. Then $\bar \mata_0$  is a rank two matrix  with the two non zero eigenvalues  $\lambda_1=a+b$ and $\lambda_2=a-b$. The eigenvector $\uu_1$ corresponding to the eigenvalue $a+b$ has coordinates 
 $$ \uu_1(i) = \frac{1}{\sqrt{2n}}, \text{ for all } i \in V$$
 and eigenvector $\uu_2$ corresponding to the eigenvalue $a-b$ has coordinates
 $$  \uu_2(i) = \left \{ \begin{array}{cc}
 \frac{1}{\sqrt{2n}} & \text{ if } i \in V_1\\
 \frac{-1}{\sqrt{2n}} & \text{ if } i \in V_2.
 \end{array} \right.
 $$

\begin{figure}[ht]
\vskip 0.2in
\fbox{
\begin{minipage}{6in}
{\bf Spectral Partition.} 
\begin{enumerate}
\item Input the adjacency matrix $\mata_0,d:=a+b$. 
	\item  Zero out all the rows and columns of $\AA_0$ corresponding to vertices whose degree is bigger than $20 d$, to obtain the matrix $\AA$. 
	\item Find the eigenspace $W$ corresponding to  the top two eigenvalues of $\AA$.
	\item Compute  $\vv_1$,  the projection of all-ones vector on to $W$
	\item Let $\vv_2$ be the unit vector in $W$ perpendicular to $\vv_1$.
	\item Sort the vertices according to their values in $\vv_2$, and let $V'_1 \subset V$ be the top $n$ vertices, and $V'_2 \subset V$ be the remaining $n$ vertices 
	\item Output $(V'_1,V'_2)$.
\end{enumerate}
\end{minipage}
}
\caption{Spectral Partition}
\label{fig:algo1}
\end{figure}

 Notice that  the second eigenvector of $\bar{\mata}_0$ identifies  the partition. We would like to use  the second eigenvector of $\mata_0$ 
to approximately identify the partition. Since $\mata_0 = \bar{\mata}_0 + \mate_0,$ perturbation theory tells us that we get  a good approximation  if $\nbr{\mate_0}$ is sufficiently small. 
However, with probability $1-o(1) $, the norm of  ${\mate_0}$ is rather large (even larger than the norm of the main term).  
In order to handle this problem, we   modify  ${\mate_0 }$ using the auxiliary deletion, at the cost of losing a few large degree vertices. 

 Let $\bar{\mata},\mata,\mate$ be the matrices obtained from $\bar{\mata}_0,\mata_0,\mate_0$ after the deletion, respectively. Let $\Delta \defeq \bar{\mata} - \bar{\mata}_0$; we have  
\begin{align*}
\mata &= \bar{\mata} + \mate \\
&= \bar{\mata}_0 + \Delta + \mate. 
\end{align*}

The key observation is that  $\nbr{\mate} $ is significantly smaller than  $\nbr{\mate_0}$.  In the next section we will show that $\nbr{\mate}=\Oh(\sqrt{d})$, with probability $1-o(1) $, while $\nbr{\mate_0}$ is $\Theta (\sqrt{\frac{\log n}{\log \log n}})$, with probability $1-o(1)$. Furthermore,  we could show that $\nbr{\Delta}$ is only $\Oh(1)$  with probability $1-o(1) $. Therefore, if the second eigenvalue gap for the matrix $\mata_0$ is greater than $C\sqrt{d}$, for some large enough constant $C$, 
then  Davis-Kahan $\sin  \Theta$ theorem would allow us to bound the angle between the second eigenvector of $\bar{\mata}_0$ and $\mata$ by an arbitrarily small constant. 
This will, in turn, enable us to recover a large portion of the blocks, proving the following statement 

\begin{theorem} \label{step1} 
\label{thm:main} There are constants $C_0$ and $C_1$ such that the following holds. For any constants $a> b >C_0$ and $\gamma  > 0$ satisfying  $ \frac{(a-b)^2}{a+b}  \geq C_1  \frac{1}{\gamma^2}  $, 
 then with probability $1-o(1)$, {\bf Spectral Partition} outputs a $\gamma$-correct partition. \end{theorem}

\begin{remark}
The parameter  $d:= a+ b$  can be estimated very efficiently from the adjacency matrix $\mata$. We take this as input for a simpler exposition. 
\end{remark}

\begin{figure}[ht]
\vskip 0.2in
\fbox{
\begin{minipage}{6in}
{\bf Partition } 
\begin{enumerate}
\item Input the adjacency matrix $\mata_0,d:= a+b$. 
\item Randomly color the edges with Red and  Blue with equal probability. 
	\item Run {\bf Spectral Partition}  on Red graph, outputting $V_1', V_2'$.
	\item Run {\bf Correction} on the Blue graph.
	\item Output the corrected sets   $V_1^{'}, V_2^{'} $.

\end{enumerate}
\end{minipage}
}
\caption{Partition}
\label{fig:algo1}
\end{figure}

Step 2 is a further correction that gives us the optimal (logarithmic)  dependence between $\gamma$ and $\frac{(a-b)^2 }{a+b}$. The idea here is to use the degree sequence to correct the mislabeled vertices.  
Consider a mislabeled  vertex  $u \in V_1' \cap V_2 $. 
As $u \in V_2$,  we expect $u$ to have $b$ neighbors in $V_1$ and $a$ neighbors in $V_2$. Assume that {\bf Spectral Partition } output $V_1', V_2'$ where 
$|V_1 \backslash V_i' | \le 0.1 n$, we expect $u$ to have at most $0.9b +0.1a$ neighbors in 
$V_1'$ and at least $0.1 b + 0.9a$ neighbors in $V_2'$. As 

$ 0.1 b  + 0.9a >  \frac{a+b} {2} >  0.9b   + 0.1 a, $  we can correctly reclassify $u$ by thresholding. 
There are, however, few problems with this argument. First, everything is in expectation. This turns out to be a minor problem; we can use a  large deviation result to  show that a majority of 
mislabeled vertices can be detected this way. As a matter of fact, the desired logarithmic dependence is achieved  at this step, thanks to the exponential probability bound in the large deviation result. 

The more serious problem is the lack of independence. Once {\bf Spectral Partition}  has  run, the neighbors of $u$ are no longer random. We can avoid this problem using a splitting trick as given in {\bf Partition}. We sample randomly 
half of the edges of the input graph and used the graph formed by them in {\bf Spectral Partition}.  After receiving the first partition, we use the other (random) half of the edges for correction. This doesn't make the two steps completely independent, but we can still prove the stated result.


The sub-routine {\bf Correction} is as follows: 

\begin{figure}[h]
\vskip 0.2in
\fbox{
\begin{minipage}{6in}
{\bf Correction.} 
\begin{enumerate}
\item Input: a partition $V_1', V_2'$ and a  Blue graph on $V_1' \cup V_2' $.
\item For any $u \in V_1'$, label $u$ {\it  bad} if the number of neighbors of $u$ in $V_2'$ is at least $\frac{a+b} {4} $ and {\it good} otherwise. 
\item Do the same for any $v \in V_2'$.
\item Correct $V_i'$ be deleting its bad vertices and adding the bad vertices from $V_{3-i} '$.
	\end{enumerate}
\end{minipage}
}
\caption{Correction}
\label{fig:algoCorrection}
\end{figure}

\begin{figure}[H]
\vspace{-0.5in}
    \centering
    \includegraphics[scale=0.33]{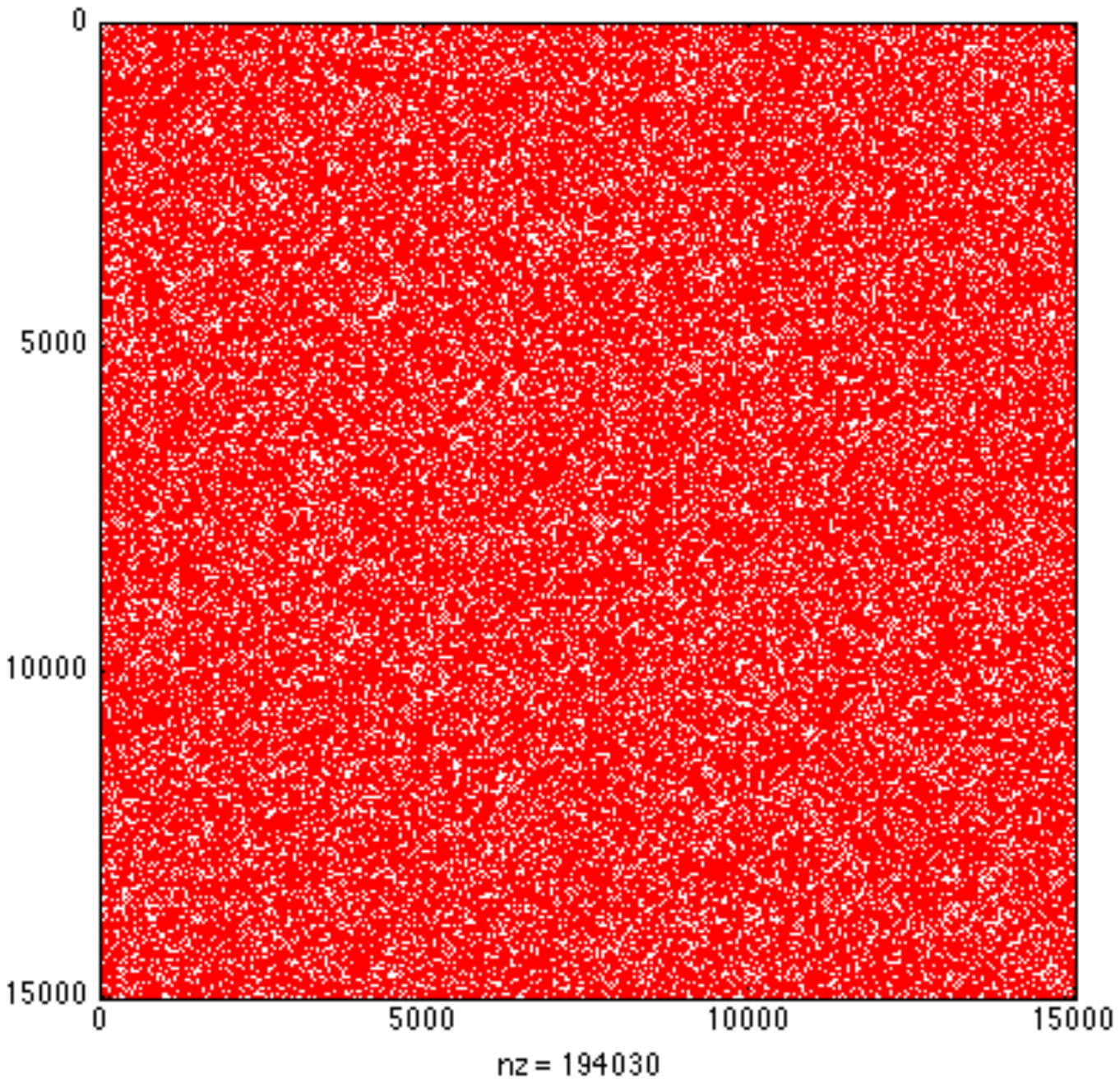}
\includegraphics[scale=0.33]{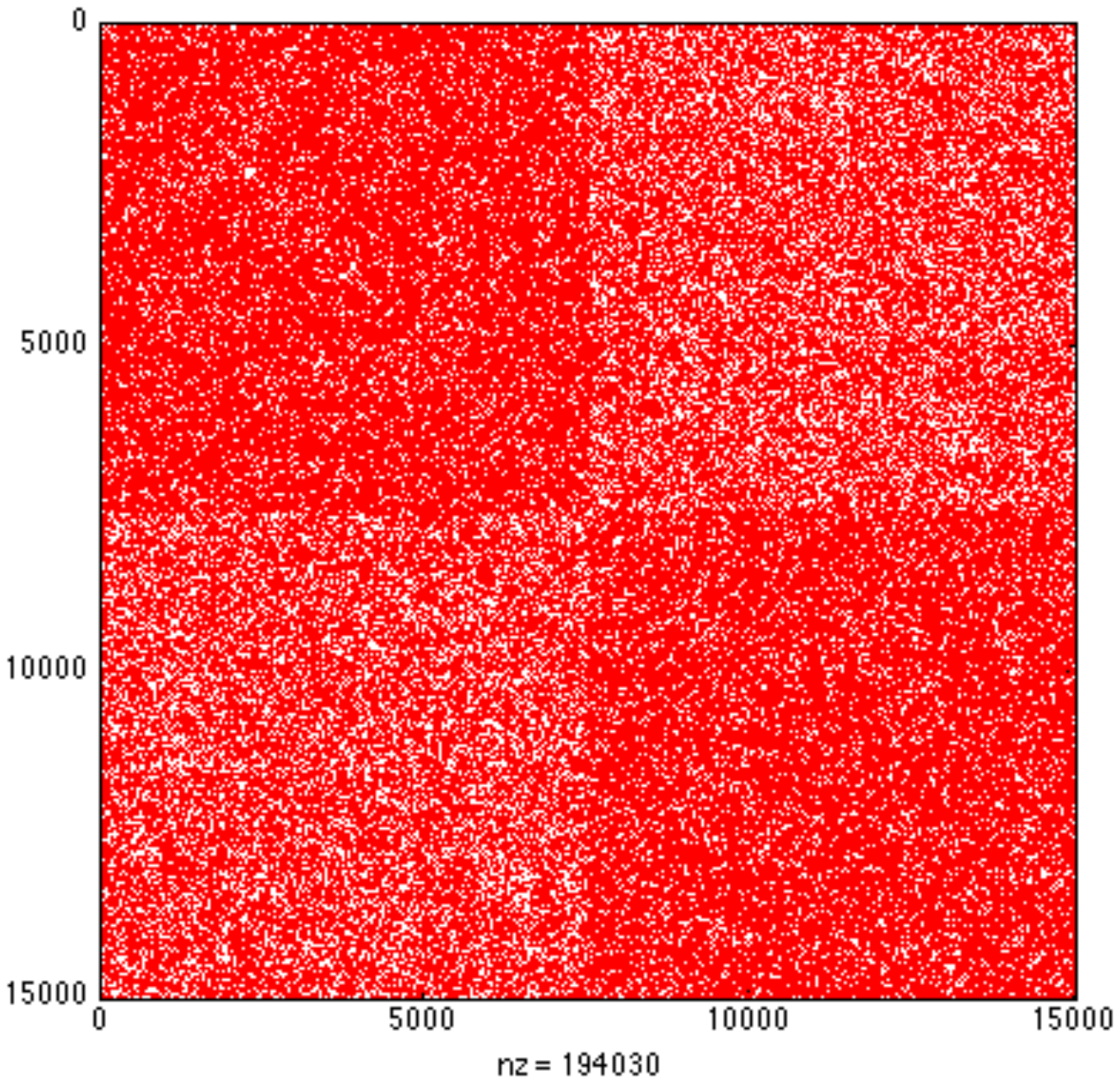}
\vspace{-1in}
\caption{On the left is the density plot of the input (unclustered) matrix with parameters $n=7500,a=10,b=3$ and on the right is the density plot of the permuted matrix after running the algorithm described above. This took less than 3secs in Matlab running on a 2009 MacPro.}
    \label{fig:experiment_image}
\end{figure}

Figure~\ref{fig:experiment_image} is the density plot of the matrix before and after clustering according to the algorithm described above. We can prove

\begin{lemma}
\label{lem:correction}
Given a $0.1$-correct partition $V_1', V_2'$ and a  Blue graph on $V_1' \cup V_2' $ as input to the subroutine {\bf Correction} given in figure~\ref{fig:algoCorrection}, we get a $\gamma$-correct partition with $\gamma = 2 \exp( - 0.072 \frac{(a-b)^2 }{a+b} ).$
\end{lemma}

\section{First step: Proof of Theorem~\ref{thm:main}}
 \label{sec:Proof}
 
 We now turn to the details of the proof. 
  Using the notation in the previous section, we let 
   $W$ be the two dimensional eigenspace corresponding to the top two eigenvalues of $\AA$ and $\bar{W}$ be the corresponding space of $\bar{A}$. For any two vector subspaces $W_1,W_2$ of same dimension, we use the usual convention $\sin \angle (W_1,W_2) := \nbr{P_{W_1} - P_{W_2}}$, where $P_{W_i}$ is the orthogonal projection onto $W_i$. 
   The proof has two main steps: 
   
 \begin{enumerate}
 \item {\it Bounding the angle}: We show that  $\sin \angle (W,\bar{W})$ is small, under the conditions of the theorem. 
 \item {\it Recovering the partition}: If $\sin \angle (W,\bar{W})$ is small, we  find an approximate partition which can then improved to find an optimal one.
 \end{enumerate} 
 
 \subsection{Bounding the angle}
 
 For the first part, recall that $\AA = \bar{\AA}_0 + \Delta + \mate$. We first prove that $\nbr{\Delta}$ and $\nbr{\mate}$ are small with probability $1-o(1) $.
 Bounding $\nbr{\Delta}$ is easy as it will be sufficient to bound the number of vertices of high degrees. We need the following

 \begin{lemma}
\label{thm:degreebound}
There exist  a  constant $d_0$ such that if $d := a+ b \geq d_0$, then with probability $1-\exp \left(-\Omega ( a^{-2} n ) \right)$  not more than $a^{-3}n $  vertices have degree  $\geq 20d$.
\end{lemma}

Note that the proof of the above lemma and other missing proofs in this subsection appear in appendix \ref{app:matrixNorm}. If there are at most $a^{-3} n$ vertices with degree $\ge 20d$, then by definition, $\Delta$ has at most  $2a^{-3} n^2$ non-zero entries, and the 
magnitude of each entry is bounded by $\frac{a}{n}$. Therefore, its  Hilbert-Schmidt norm is bounded by $\nbr{\Delta}_{HS} \leq \sqrt{2}a^{-1/2} $. 

\begin{corollary}
\label{thm:deltabound}
For $d_0$ sufficiently large,  with  probability $1 - \exp( -\Omega (a^{-3} n )) $,  $\nbr{\Delta}  \leq 1$.
\end{corollary}

Now we address the harder task of bounding $\| E \|$. Here is the key lemma

 \begin{lemma}
 \label{cor:sparseNorm}
 Suppose $M$ is random symmetric matrix with zero on the diagonal whose entries above the diagonal are independent with the following distribution
 
  $$M_{ij}=\left \{ 
\begin{array}{ccc}
1-p_{ij} & \text{ w.p. } & p_{ij}\\
-p_{ij} & \text{ w.p. } & 1-p_{ij} \end{array} \right. .$$ 

Let $\sigma$ be a quantity such that   $p_{ij}  \leq \sigma^2$ and  $M_1$ be the matrix obtained from $M$ by zeroing out all the rows and columns having 
 more than $20 \sigma^2 n$ positive entries. Then with probability $1-o(1)$, $\nbr{M_1} \leq C \sigma \sqrt{n}$ for some constant $C >0$. 
 \end{lemma}

 Lemma  \ref{cor:sparseNorm} implies 
  
 \begin{corollary} 
 \label{lem:normE}
  There exist constants $C_0, C$ such that if $a>b \geq C_0$, and $\mate$ is obtained as described before, then we have, $$\nbr{\mate} \leq C \sqrt{d}$$ with probability $1-o(1) $.
\end{corollary}   

Now, let  $\bar{\vv_1}, \bar{\vv}_2$ be eigenvectors of $\bar{\mata}_0$ corresponding to the largest  two eigenvalues $\lambda_1 \ge \lambda_2$ $\vv_1, \vv_2$ be eigenvectors of $\mata = \bar{\mata}_0 + \Delta + \mate$ corresponding to the largest two  eigenvalues.  Further, $\bar{W}:=\text{Span}\{\bar{\vv_1}, \bar{\vv}_2\}$ and $W:=\text{Span}\{\vv_1,\vv_2\}$.

\begin{lemma}
\label{lem:VW_angle}
For any constant $ c <1$,  we can choose constants $C_2$ and $C_3$ such that  such that if 
$a-b \geq C_2 \sqrt{a+b} = C_2 \sqrt{d}$ and
 $a \geq C_3$
then,  $\sin(\angle \bar{W},W) \leq c<1$ with probability $1-o(1) $.
\end{lemma}

\begin{Proofof}{Lemma~\ref{lem:VW_angle}}
Let  $C_3$ be a constant such that if $a  \geq C_3$, then theorem~\ref{thm:deltabound} holds giving us $\nbr{\Delta} \leq 1$. From lemma~\ref{lem:normE} we have that  $\nbr{E} \leq C\sqrt{d}$.  The lemma then follows from the Davis-Kahan \cite{Davis} \cite{bhatia97}  bound for matrices $\bar{\mata}_0$ and $A$, which gives  $ \sin(\angle W,\bar{W}) \leq \frac{\nbr{E + \Delta}}{\lambda_2}$. Therefore, the lemma follows by choosing $C_1$ big enough.
\end{Proofof}

\subsection{Recovery}
Given a subspace $W$ satisfying $\sin(\angle \bar{W},W) \leq c<1/16$, we can recover a big portion of the vertices. We prove  (in appendix \ref{app:Recovery}) that 
\begin{lemma}
Given a subspace $W$ satisfying $\sin(\angle \bar{W},W) \leq c<1/16$, we can recover a $8c/3$-correct partition. 
\end{lemma}
Once we have an approximate partition, we can use the Blue edges to boost it in the {\bf Correction} step. We prove (in appendix \ref{app:correction})
\begin{lemma}
\label{lem:2BlockCorrection}
Given a $0.1$ correct partition $V_1',V_2'$ as input to the {\bf Correction} routine in figure \ref{fig:algoCorrection}, the algorithm outputs a $\gamma$ correction partition with $\gamma  = 2 \exp( - 0.072 \frac{(a-b)^2 }{a+b} ).$
\end{lemma}

\section{Multiple communities}
\label{sec:MultComm}

\subsection{Overview}

Let us start with the algorithm, which 
(compared to the algorithm for the case of 2 blocks) has an additional step of random splitting. This additional step is needed in order to  recover the partitions.  We will start by computing an approximation of the space spanned by the first $k$ eigenvectors of the hidden matrix. However, when $k >2$, it is not obvious how to approximate the eigenvectors themselves. To handle this problem, we need a new  argument that requires this extra step. 
  
  \begin{figure}[h]
\vskip 0.2in
\fbox{
\begin{minipage}{6in}
{\bf Partition } 
\begin{enumerate}
\item Input the adjacency matrix $\mata_0,a,b$. 
\item Randomly color the edges with Red and  Blue with equal probability. 
\item Randomly partition $V$ into two subsets $Y$ and $Z$. Let $B$ be the adjacency matrix of the bipartite graph between $Y$ and $Z$ consisting only of the Red edges, with rows indexed by $Z$ and the columns indexed by $Y$.

	\item Run {\bf Spectral Partition} (figure \ref{fig:algoSpectral_multi})  on matrix $B$, and get  $U_1', U_2',...,U_k'$ as output. This part uses only the Red edges that go between vertices in $Y$ and $Z$ and outputs an approximation to the clustering in $Z = U_1 \cup ... \cup U_k$. Here, $U_i := V_i \cap Z$.
	\item Run {\bf Correction} (figure \ref{fig:algoCorrect_multi}) on the Red graph. This procedure only uses the Red edges that are internal to $Z$ and improves the clustering in $Z$.
         \item Run {\bf Merging} (figure \ref{fig:algoMerge_multi}) on the Blue graph. This part uses only the Blue edges that go between vertices in $Y$ and $Z$ and assigns the vertices in $Y$ to appropriate cluster.

\end{enumerate}
\end{minipage}
}
\caption{Partition}
\label{fig:algoOverview_multi}
\end{figure}

    Since we use different set of edges for each step, we have independence across the steps. 
  
  \subsection{Details}
  
Step 1 is a spectral algorithm on a portion of the adjacency matrix $A_0$ as given in figure \ref{fig:algoSpectral_multi}. 
\begin{figure}[h]
\vskip 0.2in
\fbox{
\begin{minipage}{6in}

{\bf Spectral Partition.} 

\label{line:initZ}
\begin{enumerate}
	\item Input  $B$ (a matrix of dimension $|Z| \times |Y|$ ), $a$, $b$ and $k$.
	\item Let $Y_1$ be a random subset of $Y$ by selecting each element with probability $\frac{1}{2}$ independently and let $A_1,A_2$ be the sub matrix of $B$ formed by the columns indexed by $Y_1,Y_2:= Y \backslash Y_1$, respectively.
	\item Let $d:={a+(k-1)b}$. Zero out all the rows and columns of $A_1$ corresponding to vertices whose degree is bigger than $20 d$, to obtain the matrix $A$. 
	\item Find the space spanned by $k$ left singular vectors of $A$, say $W$
	\item Let $\aa_1,...,\aa_m$ be  some $m=2 \log n$ random columns of $A_2$. For each $i$, project $\aa_i - \aa$ onto  $W$, where $\aa(j) = \frac{a+b}{2n}$ for all $j$ is a constant vector.
	\item For each projected vector, identify the top (in value) $n/2k$ coordinates. Of the $2 \log n$ sets so obtained, discard half of the sets with the lowest Blue edge density in them.
	\item Of the remaining subsets, identify some $k$ subsets $U_1',...,U_k'$ such that $|U_i' \cap U_j'| < 0.2n/2k$, for $i \neq j$. 
	\item Output $U_1',...,U_k'$.
	\end{enumerate}
	
\end{minipage}
}
\caption{Spectral Partition}
\label{fig:algoSpectral_multi}
\end{figure}
This will enable us to recover a large portion of the blocks $Z \cap V_1,...,Z \cap V_k$. We will prove the following statement (appendix \ref{app:multSpectral})

   \begin{theorem} 
\label{them:multComm2}
 There exists constants $C_1,C_2$ such that  for any fixed integer $k$ the following holds.
 
\begin{enumerate}
\item $a > b \geq C_1$ 
\item $ \frac{(a-b)^2} {a}  \geq C_2 k^2 \frac{1}{\gamma} $, and  
\end{enumerate} then we can find a $\gamma$-correct partition $U_1',...,U_k'$ of $Z$ with high probability  using a simple spectral algorithm. 
\end{theorem}

 Step 2 (figure \ref{fig:algoCorrect_multi})  is a further correction that gives us the optimal (logarithmic)  dependence between $\gamma$ and $\frac{(a-b)^2 }{a+b}$. The idea here is to use the degree sequence to correct the mislabeled vertices in Z.  
Consider a mislabeled  vertex  $u \in Z \cap V_1 $. 
As $u \in Z \cap V_1$,  we expect $u$ to have $a/4$ Red neighbors in $Z \cap V_1$ and $b/4$ Red neighbors in $Z \cap V_i$  for all $i \neq 1$. Assume that {\bf Spectral Partition } output $U_1', ...,U_k'$ where 
$|U_1 \backslash U_1' | \le .1 n/2k$, we expect $u$ to have at most $0.9b/4k +0.1a/4k$ Red neighbors in 
$U_i'$ and at least $0.1 b/4k + 0.9a/4k$ Red neighbors in $U_1'$. As 

$$ 0.1 b/8k  + 0.9a/8k >  \frac{a+b} {8k} >  0.9b/8k   + 0.1 a/8k $$  we can correctly reclassify $u$ by thresholding.  We can prove (appendix \ref{app:multCorrection})
\begin{lemma}
\label{lem:correction_multi}
Given a $0.1$ correction partition of $Z = (Z \cap V_1) \cup ... \cup (Z \cap V_k)$ and the Red graph over $Z$, the sub-routine {\bf Correction} given in figure \ref{fig:algoCorrect_multi} computes a $\gamma$ correct partition with $\gamma = 2k \exp( - 0.04 \frac{(a-b)^2 }{k(a+b)} ) $.
\end{lemma}

\begin{figure}[h]
\vskip 0.2in
\fbox{
\begin{minipage}{6in}
{\bf Correction.} 
\begin{enumerate}
\item Input: A collection of subsets  $U_1',..., U_k' \subset Z$ and a graph on $Z$.
\item For every $u \in Z$, if $i \in \{ 1,2,...,k \}$ is such that $u$ has maximum neighbors in $U_i'$, then add $u$  to $U_i''$. Break ties arbitrarily.
\item Output $U_1'',...,U_k''$.
\end{enumerate}
\end{minipage}
}
\caption{Correction}
\label{fig:algoCorrect_multi}
\end{figure}

   Step 3 is to use the clustering information of vertices in $Z$ to label the vertices in  $Y$, and is similar to step 2. We prove (appendix \ref{app:multMerge})

 \begin{lemma}
 \label{lem:merge_multi}
Given a $0.1$ correction partition of $Z = (Z \cap V_1) \cup ... \cup (Z \cap V_k)$ and the Blue graph over $Y \cup Z$, the sub-routine {\bf Merge} is given in (figure \ref{fig:algoMerge_multi}) computes a $\gamma$ correct partition with $\gamma =  2k \exp( - 0.0324 \frac{(a-b)^2 }{k(a+b)} )$.
\end{lemma}

\begin{figure}[ht]
\vskip 0.2in
\fbox{
\begin{minipage}{6in}
{\bf Merging.} 
\begin{enumerate}
\item Input: A partition $U_1',..., U_k'$  of $(Z \cap V_1) \cup (Z \cap V_2) \cup ... \cup (Z \cap V_k)$ and a graph between vertices $Y$ and $Z$.
\item For all  $u \in Y$, label $u$  with {\bf `i'} if the number of neighbors of $u$ in $U_i'$ is at least $\frac{a+b} {8} $. Label the conflicts arbitrarily.
\item Output the label classes as the clusters $V_1^{'},...,V_k^{'}$.
\end{enumerate}
\end{minipage}
}
\caption{Merge}
\label{fig:algoMerge_multi}
\end{figure}

Combining lemmas $\ref{lem:correction_multi},\ref{lem:merge_multi}$, we get the stated result.

\begin{figure}[H]
\vspace{-1in}
    \centering
    \includegraphics[scale=0.33]{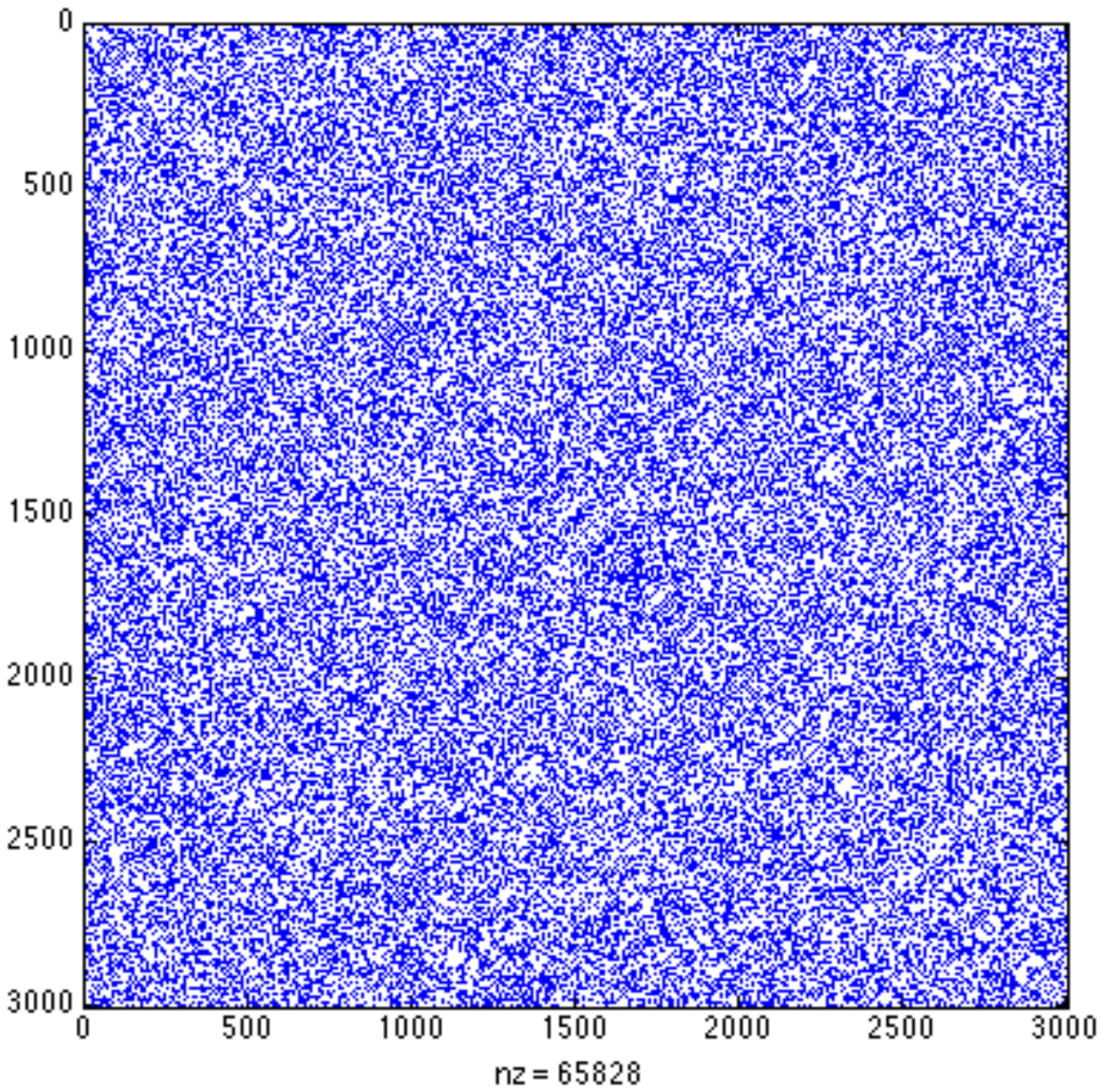}
\includegraphics[scale=0.33]{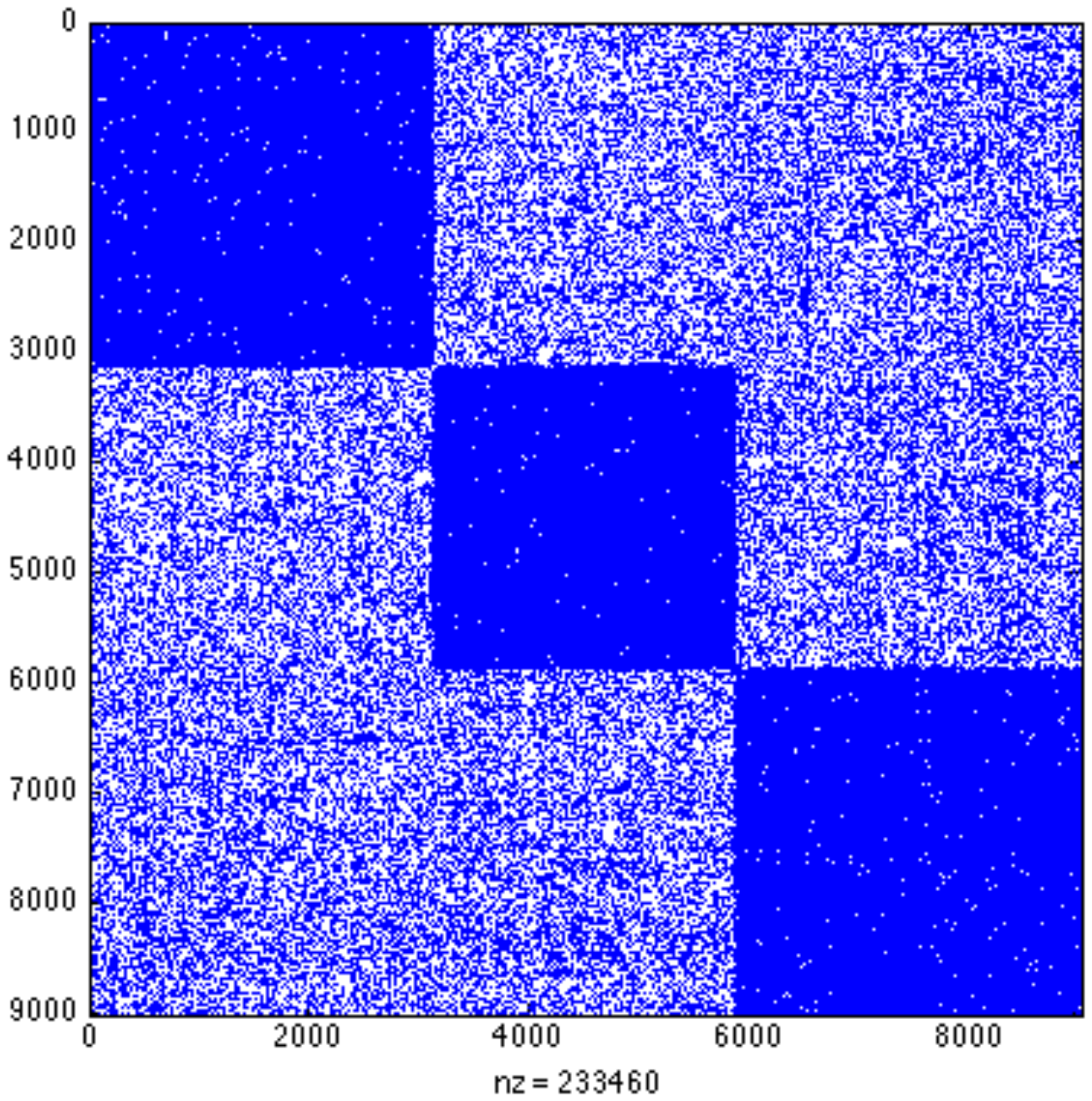}
\vspace{-1in}
\caption{On the left is the density plot of the input (unclustered) matrix with parameters $n=3000,a=22,b=2$ and on the right is the density plot of the permuted matrix after running the algorithm described above. This took less than 1sec in Matlab running on a 2009 MacPro.} 
    \label{fig:Image_kblocks}
\end{figure}

\section{Censor Block Model}
\label{sec:CBM}

We first introduce some notations so as to write this problem in a way similar to the other problems in this paper. To simplify the analysis, we make the following assumptions. We assume that there are $|V|=2n$ vertices, with exactly $n$ of them labeled $1$, and the rest labeled $0$. As in \cite{Abbe}, we assume that $G \in G_{2n,p}$ is a graph generated from the Erdos-Renyi model with edge probability $p$. Since any edge $(i,j)$ appears with probability $p$, and that $\zz_e \sim \text{Bernoulli}(\epsilon)$, we have $$y_{i,j} = \left \{
 \begin{array}{ccc} 
 x_i \oplus x_j & w.p. & p(1 - \epsilon) \\
 x_i \oplus x_j \oplus 1 & w.p. & p \epsilon \\
0 & w.p. & 1-p 
\end{array} \right. .$$
For any $i,j \in V$, let us write $w_{ij} := x_i \oplus x_j $, and $W := (w_{ij})_{ij}$ the associated $2n \times 2n$ matrix. 
\begin{figure}[h]
\vskip 0.2in
\fbox{
\begin{minipage}{6in}
{\bf Spectral Partition II.} 
\begin{enumerate}
\item Input the adjacency matrix $Y, p$. 
	\item Zero out all the rows and columns of $Y$ corresponding to vertices whose degree is bigger than $20 pn$, to obtain the matrix $Y_0$. 
	\item Find the eigenspace $U$ corresponding to  the top two eigenvalues of $Y_0$.
	\item Compute  $\vv_1$,  the projection of all-ones vector on to $U$
	\item Let $\vv_2$ be the unit vector in $W$ perpendicular to $\vv_1$.
	\item Sort the vertices according to their values in $\vv_2$, and let $V'_1 \subset V$ be the top $n$ vertices, and $V'_2 \subset V$ be the remaining $n$ vertices 
	\item Output $(V'_1,V'_2)$.
\end{enumerate}
\end{minipage}
}
\caption{Algorithm 3}
\label{fig:algo3}
\end{figure}

We note that 
$\bar{y}_{i,j} := \E (y_{i,j}) = p \epsilon + p(1 - 2\epsilon)w_{i,j}.$
Therefore, we can write $y_{i,j} = \bar{y}_{i,j} + \zeta_{i,j}$, where $\zeta_{i,j}s$ are mean zero random variables satisfying $\text{Var}(\zeta_{i,j}) \leq p.$ 
First we note that we can recover the two communities from the eigenvectors of the $2n \times 2n$ matrix $\bar{Y} := (\bar{y}_{i,j}) = p\epsilon I + p(1 - 2\epsilon) W .$  $\bar{Y}$ is a rank $2$ matrix with eigenvalues $pn$ and $p(1 - 2\epsilon)n$, with the corresponding eigenvectors $\vv_1 = (1,1,....,1)$ and $\vv_2=(1,...,1,-1,...,-1).$ if we can find $\vv_2$, we can identify the two blocks. Let $Y = (y_{i,j})$ and $E = (\zeta_{i,j})$ be $2n \times 2n$ matrices. Algorithm \ref{fig:algo3} (which is essentially same as algorithm \ref{fig:algo1}) which takes as input the adjacency matrix $Y$ and the edge probability $p$ achieves this when $np \geq \frac{C_2}{(1-2\epsilon)^2}.$  More detail appears in appendix \ref{app:CBM}.


\bibliographystyle{alpha}
\bibliography{references}

\appendix

\section{Two communities}

\subsection{ Bounding $\nbr{\Delta} and \nbr{E}$}
\label{app:matrixNorm}
\begin{Proofof}{Lemma~\ref{thm:degreebound}} One can prove Lemma \ref{thm:degreebound} using a standard argument from random graph theory. Consider a set of vertices $X \subset V$ of size $|X|= c n$, where $ c <1$ is a constant. We first bound the probability that all the vertices in this set have degree greater than $20d$.

   Let us denote the set of edges on  $X$ by $E(X)$ and the set of edges with exactly one end point in $X$ by $E(X,X^c)$.  
   If each degree in $X$ is at least $20d$, then a quick consideration reveals that  either $|E(X)| \geq 2cnd$ or $|E(X,X^c)| \geq 8cnd$. The expected number of edges $\mu_{E(X)} := \E(|E(X)|)$ satisfies $$ 0.25 (cn)^2 \frac{a}{n} \leq \mu_{E(X)} \leq 0.5 (cn)^2 \frac{a}{n} .$$ Let $\delta_1 := \frac{2}{c} \leq \frac{2cnd}{\mu_{E(X)}}$, then Chernoff bound (see \cite{alon2004} for example) gives 
 \begin{align*}
 \mathbb{P}(|E(X)| \geq cnd) &\leq \left( \frac{\exp(\delta_1-1)}{\delta_1^{\delta_1}} \right)^{\mu_{E(X)}}\\
 & \leq \exp \left( \left(  \frac{2}{c}-1 - \frac{2}{c} \log \left( \frac{2}{c} \right) \right) 0.25 (cn)^2 \frac{a}{n}  \right)\\
  & \leq \exp \left(   - \frac{1}{c} \log \left( \frac{1}{c}  \right) 0.25 (cn)^2 \frac{a}{n}  \right)  \text{ (for small enough } c) \\
 & = \exp \left( -0.25  \log \left( \frac{1}{c} \right)acn  \right).
 \end{align*}
 Similarly, the expected number of edges $\mu_{E(X,X^c)}$ in $E(X,X^c)$ satisfies $$ c(1-c)n^2 \frac{a}{n} \leq \mu_{E(X,X^c)} \leq  c(2-c)n^2 \frac{a}{n}.$$ Let $\delta_2 :=4 \leq \frac{8cnd}{\mu_{E(X,X^c)}}$, then by Chernoff  bound 
  \begin{align*}
 \mathbb{P}(|E(X,X^c)| \geq 8cnd) &\leq \left( \frac{\exp(\delta_2-1)}{\delta_2^{\delta_2}} \right)^{\mu_{E(X,X^c)}}\\
 & \leq \exp \left(- c (2-c) an \right).\\
 \end{align*}
 Now, if we substitute $c=a^{-3}$ in the above bounds, we get 
 \begin{align*}
 \mathbb{P}(|E(X)| \geq 2cnd) & \leq \exp \left( - 0.75 \log(a) a^{-2} n  \right) \\
 \mathbb{P}(|E(X,X^c)| \geq 8cnd) &\leq \exp \left(- a^{-2}n \right).
\end{align*}

 There are at most $${2n \choose cn} \leq \exp \left(-c \left( \log(\frac{c}{2})-1 \right)n \right)$$  subsets $X$ of size $|X|=cn$. Substituting $c=a^{-3}$ again, we get
 
 \begin{align*}
 {2n \choose cn} &\leq \exp \left(4 a^{-3} \log(a) n \right).
 \end{align*}
The claim follows from the union bound. 
 \end{Proofof}

\begin{Proofof}{Lemma \ref{cor:sparseNorm}}
We start by proving  a simpler result.

  \begin{lemma}
  \label{thm:denseNorm}
 Let  $M$ be  random symmetric matrix of size $n$  with zero diagonal  whose entries above the diagonal are independent with the following distribution
 
   $$M_{ij}=\left \{ 
\begin{array}{ccc}
1-p_{ij} & \text{ w.p. } & p_{ij}\\
-p_{ij} & \text{ w.p. } & 1-p_{ij} \end{array} \right. .$$ 

Let  $\sigma^2 \geq C_1 \frac{\log n}{n}$ be  a quantity such that $p_{ij}  \leq \sigma^2$ for all $i,j$, where $C_1$ is a constant.  Then 
with probability $1-o(1)$, $\nbr{M} \leq C_2 \sigma \sqrt{n}$ for some constant $C_2 >0$. 

 \end{lemma}
 

Let us address Lemma \ref{thm:denseNorm}. 
A weaker bound  $C \sigma \sqrt{n\log n}$  follows easily from Alshwede-Winter type matrix concentration results (see \cite{Tropp}). 
To prove the claimed bound, we need to be more careful and 
follow the $\epsilon$-net  approach by Kahn and Szemeredi  for random regular graphs in \cite{Friedman:1989:SER:73007.73063} (see also \cite{Alon94aspectral, RSA:RSA20089}). 

Consider  a $\frac{1}{2}$-net $\mathcal{N}$ of the unit sphere  $\mathcal{S}^n$. We can assume $| \mathcal{N}| \le 5^n$.
 It suffices to 
prove that there exists a constant $C'_2$ such that with probability $1-o(1)$,   $|\xx^T M \yy| \leq C'_2 \sigma \sqrt{n}$ for all $\xx,\yy \in \mathcal{N}$.   

For two vectors $\xx, \yy \in \mathcal{N}$, we follow an argument of Kahn and Szemer\'edi \cite{Friedman:1989:SER:73007.73063} and 
 call all pairs $(i,j)$ such that $|x_i y_j| \leq \frac{\sigma}{\sqrt n} $ {\it light}  and all remaining pairs {\it heavy} and  denote these two classes by $L$ and $H$ respectively.  We have   
 
 $$\vecx^T M \vecy = \sum_{i,j} x_i M_{ij} y_j = \sum_{L} x_i M_{i,j} y_j + \sum_{H} x_i M_{i,j} y_j . $$
We now show that with probability $1-o(1)$, the last two summands are small in absolute value. 
 
  First, let us consider the contribution of light couples.  We rewrite  $X := \sum_{L} x_i M_{i,j} y_j$ as $\sum_{(i,j) \in L, i>j} M_{i,j}a_{i,j} $, where  $$a_{i,j}=\left\{ \begin{array}{ccc}
x_{i}y_{j}+x_{j}y_{i} & \text{ if } & (i,j),(j,i)\in L\\
x_{i}y_{j} & \text{ if } & (i,j)\in L\\
x_{j}y_{i} & \text{ if } & (j,i)\in L\end{array}\right. $$ 
By the definition light  pairs, $|a_{i,j}| \leq 2 \frac{\sigma}{\sqrt n}$. Also, since $\vecx$ and $\vecy$ are unit vectors,  $\sum_{i,j}a_{i,j}^2 \leq 4$. Therefore, by Bernstein's bound (see page 36 in \cite{boucheron} for e.g.)
 
  $$\mathbb{P}(X>t) \leq \exp \left( \frac{-\frac{1}{2}t^2}{4 \sigma^2 + \frac{1}{3}2\frac{\sigma}{\sqrt{n}}t } \right ).$$

  
Set $t =10 \sigma \sqrt n$ and use the union bound (combining with the fact that the net has at most $5^n$ vectors, we can conclude that with probability 
at least $1 -\exp(-3n )$,  $|\sum_{L} x_i M_{i,j} y_j| \le 10 \sigma $.

Next we handle the heavy pairs in  $H$. Since $1 \ge \sum_H x_i^2 y_j^2 $, the definition of {\it heavy} implies that   $ \sum_{H} |x_i y_j | \leq \frac{\sqrt{n}}{\sigma}$.

  Let  $A_{i,j} := M_{i,j} + p_{i,j}$, then   $$\sum_{H} x_i M_{i,j} y_j = \sum_{H} x_i A_{i,j} y_j  -  \sum_{H} p_{ij} x_i y_j  .$$  Note that $A$ defines a graph, say $G_A$, such that $A$ is its adjacency matrix. As $p_{ij} \le  \sigma^2$, we have   $ \sum_{H} p_{ij}|x_i y_j | \leq \sigma^2 \frac{\sqrt n}{\sigma} = \sigma \sqrt{n}$. We use the following lemma to bound the first term.

  \begin{lemma}
 \label{lem:FE_heavy}
  Let $\widetilde{G}=(\widetilde{V},\widetilde{E})$ be any graph whose adjacency matrix is denoted by $\widetilde{\mata}$, and $\xx,\yy$ be any two unit vectors.  Let $\widetilde{d}$ be such that the maximum degree $\leq c_1\widetilde{d}$. Further, let $\widetilde{d}$ satisfy the property that for any two subsets of vertices $S,T \subset \widetilde{V}$ one of the following holds for some constants $c_2$ and $c_3$:
 
 \begin{align}
 \label{enum:descrep1}
  \frac{e(S,T)}{|S||T|\frac{\widetilde{d}}{n}} \leq c_2
  \end{align}
  
  \begin{align}
  \label{enum:descrep2}
 e(S,T) \log \left ( {\frac{e(S,T)}{|S||T|\frac{\widetilde{d}}{n}}} \right ) \leq c_3 |T| \log{\frac{n}{|T|}}
 \end{align}
 
 then $\sum_{H} x_i \widetilde{\mata}_{i,j} y_j \leq \max (16,8c_1, 32c_2, 32c_3) \sqrt{\widetilde{d}}$. Here $H:= \{ (i,j) | |x_i y_j| \geq \sqrt{\widetilde{d}}/n$ \}.
 \end{lemma}
 
 The proof appears in appendix \ref{sec:proofFE}. 
  
 \begin{lemma}
 \label{lem:dense_graph}
 Let $\widetilde{d} := \sigma^2 n$. Then with probability $1-o(1)$, the maximum degree in the graph $G_A$ is  $\leq 20 \widetilde{d}$  and for 
  any $S,T \subset V$ one of the conditions (\ref{enum:descrep1}) or (\ref{enum:descrep2} ) holds. 
 \end{lemma}

 The two lemmas above guarantee that with probability $1-o(1)$,  $| \sum_{H} x_i A_{i,j} y_j|  \leq C' \sigma \sqrt{n} $  for some constant $C'$.
  
 \Proof
 The bound on the maximum degree follows from the Chernoff bound.  We have that $$A_{ij}=\left \{ 
\begin{array}{ccc}
1 & \text{ w.p. } & p_{ij}\\
0 & \text{ w.p. } & 1-p_{ij} \end{array} \right. .$$ 
Consider a particular vertex $k$ and let $X = \sum_{i} A_{ik}$ be the random variable denoting the number of edges incident on it. We have that $$\mu = \E X = \sum_{i} p_{ik} \leq \sigma^2 n .$$ For any $l \geq 4$, Chernoff bound (see \cite{alon2004}) implies that  

\begin{align*}
\mathbb{P} (X > l \sigma^2 n) &\leq \exp \left( {- \frac{\sigma^2 n l \ln l }{3}}\right) \\
& \leq \exp \left (- \frac{l  \log n}{3} \right).
\end{align*}
Applying this with $l=20$, and taking a union bound over all the vertices, we can bound the maximum degree by $20 \sigma^2 n$. 
Now let $S,T \subset V$ be any two subsets. Let $X:= e(S,T)$ be the number of edges going between $S$ and $T$. We have $\E X \leq \sigma^2 |S||T|$. If $|T| \geq \frac{n}{e}$, then since the maximum degree is $\leq 20 \sigma^2 n$, we have $e(S,T) \leq  |S| 20 \sigma^2 n \leq 20e \sigma^2 |S||T| $, giving us \ref{enum:descrep1} in this case. Therefore, we can assume $|T| \leq \frac{n}{e}$. By Chernoff bound, it follows that for any $l \geq 4$, \begin{align*}
\mathbb{P} (e(S,T) > l \sigma^2 |S||T|) &\leq \exp \left( {- \frac{l \ln(l) \sigma^2 |S||T|}{3}}\right). 
\end{align*}
Let $l'$ be the smallest number such that $l' \ln(l') \geq \frac{21|T|}{\sigma^2 |S||T|} \log \left( \frac{n}{|T|} \right)$. As in \cite{RSA:RSA20089}, if we choose $l = \max (l',4)$, we can bound the above probability by $\exp \left( {- \frac{l \ln(l) \sigma^2 |S||T|}{3}}\right) {n \choose |S|} {n \choose |T|} \leq \frac{1}{n^3}$. Therefore, by the union bound we get that with probability $1-o(1) $ for all subsets $S,T$, and   $$e(S,T) \leq  \max (l',4) \sigma^2 |S||T| .$$  This implies that one of the conditions \ref{enum:descrep1} or \ref{enum:descrep2} holds with probability $1-o(1) $. 

 \QED

 \begin{Proofof}{Lemma \ref{cor:sparseNorm}}Now we are ready to prove Lemma \ref{cor:sparseNorm} by modifying the previous proof. 
 We again handle the light couples and the heavy couples separately, but need to  make a modification to the argument for the light couples. 
 
 Since we zero out some rows and columns of $M$ to obtain $M_1$, we first bound the norm of the matrix  $M_0$, obtained from $M$ by zeroing out a  set $S$  of rows and the corresponding columns. Next, we  take a union bound over all choices of $S$. For a fixed $S$, lemma \ref{thm:denseNorm} implies that  with probability at least $1 -\exp( -3n)$, 
 for all  $\xx, \yy \in \mathcal{N}_{1/2}$,  $|\sum_{L} x_i (M_0)_{ij} y_j| \leq 10 \sigma \sqrt{n} $. Since there are at most $2^n = \exp \left(n\ln2 \right)$ choices for $S$, 
  we can apply a union bound to show that  with probability at least $ 1 - \exp (-(3 - \ln2)n)$, $ |\sum_{L} x_i (M_1)_{ij} y_j| \leq 10 \sigma \sqrt{n} $.

  The proof for the heavy couples goes through without any modifications. We just have to verify that the conditions of lemma \ref{lem:FE_heavy} are met. Firstly, the adjacency matrix $A_1$
   obtained from $M_1$  has bounded degree property by the definition of $M_1$. Now we note that only for the case of $|S| \leq |T| \geq \frac{n}{e}$ did we need that the maximum degree was bounded. So for any $|S| \leq |T| <  \frac{n}{e}$, the discrepancy properties (\ref{enum:descrep1}) or (\ref{enum:descrep2}) holds  for $A_1$, since zeroing out rows and columns can only 
   decrease the edge count across sets of vertices. In the case $|T| \geq  \frac{n}{e}$, like before we can show that (\ref{enum:descrep1}) holds for $A_1$ since the degrees are bounded.
 \end{Proofof}  
\QED

  Now,  to bound the norm of matrix $E$, we just appeal to \ref{cor:sparseNorm}. Suppose $a >b  \geq C_0$, for a large enough constant  $C_0$ to be determined later. Since $\mata = \bar{\mata}_0 + \Delta + \mate$ and we have bounded $\Delta$, it remains to bound $\nbr{\mate}$.   Note that $$(\mate_0)_{ij} =  \left \{ 
\begin{array}{ccc}
1-\frac{a}{n} & \text{ w.p. } & \frac{a}{n} \\
- \frac{a}{n} & \text{ w.p. } & 1-\frac{a}{n} 
 \end{array} \right. $$ 
 if   $i,j$  belongs to the same community and $$(\mate_0)_{ij} =  \left \{ 
 \begin{array}{ccc}
1-\frac{b}{n} & \text{ w.p. } & \frac{b}{n} \\
- \frac{b}{n} & \text{ w.p. } & 1-\frac{b}{n} 
 \end{array} \right. $$ 
  if   $i,j$  belongs to different communities. Since $a>b$, for all $i,j$ we have that $$\text{Var}((\mate_0)_{ij}) \leq \frac{a}{n} \left(1 - \frac{a}{n} \right) \leq \frac{d}{n}.$$  
 \end{Proofof}
 
 \subsection{Recovery}
\label{app:Recovery}
Now we focus on the second step in the proof, namely the recovery of the blocks once the angle condition is satisfied. 

\begin{lemma}
\label{lem:vu_angle}
If  $\sin(\angle \bar{W},W) \leq c  \leq \frac{1}{4}$, then we can find a vector $\vv \in W$ such that $\sin(\angle \vv, \bar{\vv_2}) \leq 2 \sqrt{c}$. 
\end{lemma}

\Proof
Let $P_{\bar{W}}, P_W$ be the orthogonal projection operators on to the subspaces $\bar{W},W$ respectively. From the angle bound for the subspaces, we have that $$\nbr{P_{\bar{W}} - P_W}_2 \leq c.$$ The vector we want is obtained as follows. We first project $\bar{\vv}_1$ on to $W$, and then find the unit vector orthogonal to the projection in $W$. We will now prove that the vector so obtained satisfies the bound stated in the lemma. Since $\bar{\vv_1},\bar{\vv}_2 \in \bar{W}$, we have that $\nbr{P_W \bar{\vv_i} - \bar{\vv}_i}_2 \leq c$ for $i=1,2$. Let us define $\uu_i := P_W \bar{\vv_i}$ and $\vecx_i := \uu_i - \bar{\vv}_i$ (note that $\nbr{\vecx_i} \leq c$) for $i=1,2$. We will now show that the vector $\vv \in W$ perpendicular to $\uu_1$ is close to $\bar{\vv}_2$. Let $\uu_{\perp} = \uu_2 - \frac{\uu_1^T\uu_2}{\nbr{\uu_1}^2} \uu_1$, it is then clear that $\nbr{\uu_{\perp}} \leq 1$. Note that $|\uu_1^T \uu_2| = |\bar{\vv}_1^T \vecx_2 + \bar{\vv}_2^T \vecx_1 + \vecx_1^T \vecx_2| \leq 2c + c^2$.   We have, 

\begin{align*}
\uu_{\perp}^T \bar{\vv}_2 &= \uu_2^T \bar{\vv}_2 - \frac{(\uu_1^T\uu_2) (\bar{\vv}_2^T \uu_1)}{\nbr{\uu_1}^2} \\
\left | \uu_{\perp}^T \bar{\vv}_2 \right | & \geq 1-c - \frac{(2c+c^2)c}{(1-c)^2} \\
& \geq 1 -2c.
\end{align*}
The last inequality holds when $c \leq \frac{1}{4}$. Therefore,  it holds that for a unit vector $\vv \perp \uu_1$, $$ |\vv^T \bar{\vv}_2| \geq \left | \uu_{\perp}^T \bar{\vv}_2 \right | \geq 1-2c.$$ This gives  $\sin (\angle \vv ,\bar{\vv}_2) \leq \sqrt{1 - (1-2c)^2} \leq 2 \sqrt{c} $.
\QED

\noindent Lemmas \ref{lem:VW_angle} and \ref{lem:vu_angle} together give 

\begin{corollary}
For any constant $c<1$, we can choose constants $C_2$ and $C_3$ in lemma~\ref{lem:VW_angle} and find a vector $\vv$ such that $\sin(\angle \bar{\vv}_2,\vv) \leq c<1$ with probability $1-o(1) $. 
\end{corollary}

We now can conclude the proof of our theorem using the following deterministic fact. 

\begin{lemma}
\label{lem:frac}
If $\sin(\angle \bar{\vv}_2,\vv) < c \leq 0.5$, then we can identify  at least a $(1-\frac{4}{{3}}c^2)$ fraction of vertices from each block correctly.
\end{lemma}

\begin{Proofof}{Lemma~\ref{lem:frac}}
Let us define two sets of vertices, $V'_1 = \{i | \vv(i) > 0 \}$ and $V'_2 = \{ i | \vv(i) < 0 \}$. One of the sets will have less than or equal to $\frac{n}{2}$ vertices, let us assume without loss of generality that $|V'_1| \leq \frac{n}{2}$. Writing $\vv = c_1 \bar{\vv}_2 + \err$, for a vector $\err$ perpendicular to $\bar{\vv}_2$ and  $\nbr{\err} < c$. We also have $c_1 > \sqrt{1-c^2}$. Since $\nbr{\err} < c$, not more than $\frac{c^2}{1-c^2}n$ coordinates of $\err$ can be bigger than $\frac{\sqrt{1-c^2}}{\sqrt{n}} < \frac{c_1}{\sqrt{n}} $. Since $\vv= c_1\bar{\vv}_2 + \err$ at least $1-\frac{c^2}{1-c^2}> 1-\frac{4}{3}c^2$ (since $c\leq 0.5$) fraction of vertices with $\bar{\vv}_2(i) = \frac{1}{\sqrt{n}}$ will have $\vv(i) > 0$. Therefore, we get that there are at least $(1-\frac{4}{3}c^2)n$ vertices belonging to the first block.
\end{Proofof}
\QED

 \subsection{Proof of lemma \ref{lem:correction} } 
 \label{app:correction} 

We will use the following large deviation result (see page 36 in \cite{boucheron} for e.g.) repeatedly

\begin{lemma} (Chernoff) If $X$ is a sum of $n$ iid indicator random variables with mean at most  $\rho \le 1/2$, then for any $t >0$

$$\max \{ \mathbb{P} (X \ge \E X +t) , \mathbb{P} (X \le \E X -t ) \} \le \exp \left( -\frac{t^2}{2\var X + t} \right) \le \exp \left( -\frac{t^2}{ 2 n \rho +t  }\right)  . $$ 

\end{lemma} 

In the Red graph, the edge densities are $a/2n$ and $b/2n$, respectively.  
By Theorem \ref{step1}, there is a constant $C$ such that if  $\frac{(a-b)^2 }{a+b} \ge C$  then by running {\bf Spectral Partition} on the Red graph, we obtain,  with probability $1-o(1)$ two sets $V_1'$ and $V_2'$, where 

$$|V_i \backslash V_i'| \le .1 n. $$

In the rest, we condition on this event, and  the event that the maximum Red degree of a vertex is at most $\log^2 n$, which occurs with probability $1-o(1)$.

Now we use the Blue edges. Consider $e= (u,v)$. If $e$ is not a red  edge, and $u  \in V_i, v \in V_{3-i}  $, then $e$ is a Blue edge with probability 

$$\mu: =  \frac{ b/2n }{ 1 - \frac{b}{2n }} . $$

Similarly, if $e$ is not a Red edge, and $u , v \in V_i$, then $e$ is a Blue edge with probability 

$$\tau : =  \frac{ a/2n }{ 1-  \frac{a}{2n }} . $$

Thus, for any $u \in V_i'\cap V_i$,  the number of  its  Blue neighbors in $V_{3-i} '$ is at most 

$$S(u) := \sum_{i=1}^{.9n} \xi_{i}^u   + \sum_{j=1}^{.1 n} \zeta_{ j} ^u $$

\noindent  where $\xi_{i}^u $ are iid indicator variables with mean $\mu$ and $\zeta_{j}^u$ are iid indicator variables with mean $\tau$.

Similarly, for any $u \in V_1' \cap V_2$, the number of  its  Blue neighbors in $V_2'$ is at  least 

$$S'(u) := \sum_{i=1}^{.9n -d(u) } \zeta_{i}^u   + \sum_{j=1}^{.1 n} \xi_{ j} ^u,$$ where $d(u)= \log^2 n$ is the Red degree of $u$.

After the correction sub-routine, a vertex $u $ in the (corrected) set $ V_1'$ is misclassified if  

\begin{itemize} 

\item $u \in V_1' \cap V_1 $ and $S_u  \ge \frac{a+b}{4} $. 

\item $u \in  V_1' \cap V_2$  and $S_u' \le \frac{a+b} {4}  $

\end{itemize}

Let $\rho_1, \rho_2$ be the probability of the above events. Then the number of misclassified vertices in the (corrected) set $V_1'$ is at most 

$$ M: = \sum_{ k=1}^{n}   \Gamma_k  + \sum_{l=1}^ {0.1n} \Lambda _l  $$ where $\Gamma_k$ are iid indicator random variables with mean $\rho_1$ and $\Lambda_l$ are iid indicator random variables with mean $\rho_2$.

The rest is  a simple computation.  First we use Chernoff bound to estimate $\rho_1, \rho_2$.  Consider 

$$\rho_1 := \mathbb{P} \left( S(u) \ge \frac{a+b}{4} \right). $$

By definition, we have 

\begin{equation} \label{expectation}
\begin{split}
 \E S(u) &= 0.9n \mu + 0.1n \tau \\
 &= 0.9 n(\frac{ b/2n }{ 1 - \frac{b}{2n }} ) +0.1 n(\frac{ a/2n }{ 1-  \frac{a}{2n }} ) \\
 &  =  0.9 \frac{b}{2} + 0.1 \frac{a}{2} + 0.9 \frac{b}{2} ( \frac{1}{1- b/2n } -1) + 0.1 \frac{a}{2} ( \frac{1}{1-a/2n} -1). 
 \end{split}
\end{equation} 

Set 

$$t:= \frac{a+b}{4} - \E S(u), $$ we have 

$$t = 0.2 (a-b) - 0.9 \frac{b}{2} ( \frac{1}{1- b/2n } -1) - 0.1 \frac{a}{2} ( \frac{1}{1-a/2n} -1) \ge 0.2(a-b) - 0.9 \frac{b}{2} \frac{b}{n} - 0.1 \frac{a}{2} \frac{a}{n} \ge 0.19 (a-b) , $$ for any sufficiently large $n$.

Applying Chernoff's bound, we obtain 

$$\rho_1 \le \exp( - \frac{ (0.19 (a-b) )^2 }{ 2 (0.9n \mu + .1 n \tau ) + 0.19 (a-b) } ) . $$

By \eqref{expectation}, one can show that  $2 (0.9n \mu + .1 n \tau ) + 0.19 (a-b) = 0.71 b + 0.29 a +o(1)  \le \frac{a+b}{2} $. It follows that 

$$\rho_1 \le \exp ( - 0.072 \frac{(a-b)^2 }{a+b} ). $$

By a similar argument, we obtain the  same estimate for $\rho_2$ (the contribution of the  term $d(u) \le \log^2 n $ is negligible). Thus, we can conclude that 

$$\E M \le 1.1 n \exp(- 0.072 \frac{(a-b)^2 }{a+b} ). $$

Applying Chernoff's with $t : =0.9 n  \exp(- 0.072 \frac{(a-b)^2 }{a+b} )$, we conclude that with probability $1-o(1)$

$$M  \le \E M+t  = 2 n \exp( -0.072 \frac{(a-b)^2 }{a+b} ). $$

This implies that with probability $1-o(1)$, 

$$| V_1' \backslash V_1| \le  2 n \exp( - 0.072 \frac{(a-b)^2 }{a+b} ). $$ By symmetry, the same conclusion holds for $|V_2' \backslash V_2|$.

Set

$$\gamma:  = 2 \exp( - 0.072 \frac{(a-b)^2 }{a+b} ), $$ we have, for $i=1,2$

$$|V_i \cap V_i '|  = n - |V_i \cap V_{3-i} '| = n - | V_{3-i} ' \backslash V_{3-i} |  \ge n (1- \gamma ). $$ 

This shows that the output $V_1', V_2'$ form a $\gamma$-correct partition, with $\gamma$ satisfying 

$$\frac{(a-b)^2 }{a+b} = \frac{1}{0.072} \log \frac{2}{\gamma} \approx 13.89 \log \frac{2}{\gamma} , $$ proving our claim. 

\begin{Proofof}{Corollary \ref{cor:main1}}
Notice that in the analysis of {\bf Spectral Partition}, we only require $\frac{(a-b)^2 }{a+b} \ge C$ for a sufficiently large constant $C$
(so $\gamma$ does not appear in the bound). 
In the analysis of {\bf Correction}, we require $\frac{(a-b)^2 }{a+b} \ge 13.89 \log \frac{2}{\gamma} , $ as shown above. If $\gamma < \epsilon$ for a sufficiently small $\epsilon$, this assumption implies the first. Thus, 
Corollary holds with  assumption $\frac{(a-b)^2 }{a+b} \ge 13.89 \log \frac{2}{\gamma} $.

The constant $13.89$ comes from the fact that the partition obtained from {\bf Spectral Partition} is $.1$-correct. If one improves upon $.1$, one improves $13.89$. In particular,  there is a constant $\delta$ such that if the first partition is $\delta$-correct, then one can improve $13.89$ to $8.1$ (or any constant larger  than $8$--which is the limit of the method,  for that matter). 
\end{Proofof}

\section{Multiple communities}
\label{app:multCommunities}
We say the splitting is `perfect' if we have $|Y_1 \cap V_i| = \frac{n}{4k} = |Y_2 \cap V_i|$ for $i=1,..,k.$ We will assume the splittings are perfect in the proofs for a simpler exposition. Though the splitting will almost always not be perfect, and there will just be a $o(1)$ error term that we have to carry throughout to be precise.  The bounds we give will all be still be essentially the same.

\subsection{Proof of theorem \ref{them:multComm2}} 
\label{app:multSpectral}
To analyze this algorithm, we use the machinery developed so far combined with some ideas from \cite{2014Vu}.
 We consider the stochastic block model with $k$ blocks of size $n$, where $k$ is a fixed constant as $n$ grows. This is a graph $V =V_1 \cup V_2 \cup ... \cup V_k$ where each $|V_i| =n/k$ and for $u \in V_i, v \in V_j$: $$\mathbb{P}[(u,v)\in E]=\left \{ 
\begin{array}{cc}
a/n & \text{ if } i=j\\
b/n & i \neq j \end{array} \right.$$ 
We can write, as before 
\begin{align*}
A &= \bar{A} + E   \\
&= \bar{A_1} +\Delta+ E, 
\end{align*}

 where $\bar{A}, \bar{A_1}$ are the expected matrices, and $\Delta$ is matrix containing the deleted rows and columns. Let $\bar{W}$ be the span of the $k$ left singular vectors of $\bar{A}_1$ We can bound $\nbr{\Delta} \leq 1$  by bounding the number of high degree vertices as we did before. $\mate$ is given by $$\mate_{u,v}=\left \{ 
\begin{array}{ccc}
1- \frac{a}{n} & \text{ w.p. } & \frac{a}{n}\\
-\frac{a}{n} & \text{ w.p. } & 1-\frac{a}{n} \end{array} \right.  $$
if $u,v \in V_i \cap Y_1 \text{ for some } i \in {1,..,k}$ and 
$$\mate_{u,v}=\left \{ 
\begin{array}{ccc}
1- \frac{b}{n} & \text{ w.p. } & \frac{b}{n}\\
-\frac{b}{n} & \text{ w.p. } & 1-\frac{b}{n} \end{array} \right.  $$ $ \text{ if } u \in V_i \cap Y_1 \text{ and } v \in V_j \cap Y_1 \text{ for } i\neq j .$ Since $\sigma^2 := \frac{a}{n} \geq \text{Var}(\mate_{u,v}) $, corollary \ref{cor:sparseNorm} applied to $\bar{A}_1 - A$ gives the following result.
\begin{lemma}
\label{lem:mult_normE}
There exists a constant $C$ such that $\nbr{E} \leq C \sqrt{a+b}$ with probability $1-o(1) $. 
\end{lemma}



It is not hard to show that the rank of the matrix $\bar{A_1}$ is $k$, and its least non-trivial singular value is  $\sigma_k(\bar{A}_1) = \frac{a-b}{k}.$ This fact, combined with lemma~\ref{lem:mult_normE} and an application of Davis-Kahan bound gives

\begin{lemma}
\label{lem:k_angle}
For any $c>0$, there exists constants $C_1,C_2$ such that if $(a-b) > C_1 k^2 a$ and $a > b \geq C_2$, then $\sin \angle(\bar{W},W) \leq c$ with probability $1-o(1) $. 
\end{lemma}
 
 We pick $m =  2\log n$ indices uniformly randomly from $Y_2$ and project the corresponding columns from the matrix $B$.
 Let $\bar{\aa}_{i_1},...,\bar{\aa}_{i_m}$  and $\ee_{i_1},...,\ee_{i_m}$ be the corresponding columns of $\bar{A}_1$ and $\mate$, respectively. For a subspace $W_0$, let $P_{W_0}$ be the projection on to the space $W_0$. Note that if vertex $i \in V_{n_i} \cap Y_2$, then $$\bar{\aa}_i(j) = \left \{
 \begin{array}{cc}
 \frac{a}{{n}} & \text{ if } j \in V_{n_i}\cap Z \\
  \frac{b}{{n}} & \text{ otherwise }
\end{array} \right. .$$ 

We let the vector $\bar{\aa}$ be $$\bar{\aa}(j) := \left \{
\begin{array}{cc}
 \frac{a+b}{2{n}} & \text{ if } j \in Z \\
 \end{array} \right. .$$

and  $\bb_i = \bar{\aa}_i - \bar{\aa}$.  We therefore have  $$\bb_i(j)= \left \{
 \begin{array}{cc}
 \frac{a-b}{{2n}} & \text{ if } j \in V_{n_i}\cap Z \\
 \frac{b-a}{{2n}} & \text{ otherwise }
\end{array} \right. .$$ 

Since both $\bar{\aa}_i, \bar{\aa}$ are in the column span of $\bar{A}_1$, we have for all $i$ $$\bb_i = P_{\bar{W}} \bb_i .$$

We also note that $\nbr{\bb_i} = \frac{(a-b)}{2 \sqrt{2n}}$.  Therefore, if we can recover ${\bb}_i$, we can  identify the set $V_{n_i} \cap Z$. We now argue that we can recover $\bb_i$ approximately. Since $\aa_i - \bar{\aa}= {\bb}_i + \ee_i$, we have 
\begin{align*}
P_{W} (\aa_i- \bar{\aa}) &=  P_{W} {\bb}_i+ P_{W} \ee_i \\
 &=  P_{\bar{W}}{\bb}_i + P_{{W}} \ee_i + \error_i\\
&= {\bb}_i+ P_{{W}} \ee_i + \error_i,
\end{align*}
where  $\error_i =   \left( P_{W} - P_{\bar{W}} \right) {\bb}_i.$ Since $\sin \angle(\bar{W},W) \leq \delta_1$, we have for any unit vector $\vv$, $\nbr{P_{W} \vv - P_{\bar{W}} \vv} \leq \delta_1 ,$ which in turn implies for all $i$  $$\nbr{ \error_i} \leq \delta_1 \nbr{{\bb}_i}.$$

 Therefore, it is enough to bound $\nbr{P_{{W}} \ee_i }$. We recall that $k$ is a constant that does not depend on $n$. $W$ is $k$ dimensional space giving $\mathbb{E}\nbr{P_W \ee_i}^2 \leq k \sigma^2$. By Markov's inequality, it follows that  $$\mathbb{P} (\nbr{P_{W} \ee_i} > 2 \sigma k^{1/2} ) \leq  \frac{1}{4}.$$ 
By a simple application of Chernoff bound, we have
\begin{lemma}
With probability at least $1- o(1)$, at least $m/2$ of the vectors $\ee_{i_1},...,\ee_{i_m}$ satisfy $$ \nbr{P_{W} \ee_{i_j}} < 2 \sigma k^{1/2} .$$
\end{lemma}
Let $m_1 \geq m/2$ denote the number of such vectors, hence referred to as good vectors. To avoid introducing extra notation, let us say $\ee_{i_1},..,\ee_{i_{m_1}}$ are the good vectors and the corresponding indices as good indices. Note that $\sigma \leq \frac{\sqrt{a}}{\sqrt{n}}$. For any $\delta_2 >0$, there exists a big enough constant $C_1$ such that  if $(a-b)>C_1 \sqrt{ka}$, we have that $2 \sigma k^{1/2} \leq \delta_2 \nbr{{\bb_{i_j}}}$ whenever $i_j$ is good. Therefore
 
 \begin{lemma}
  Given any $\delta >0$, there exists constants $C_1,C_2$ such that the following holds. If $(a-b)>C_1 \sqrt{ka}$ and $a \ge b \geq C_2$, then for all good indices $i_j$, it holds that $\nbr{P_W (\aa_{i_j} - \bar{\aa})-\bb_{i_j}} \leq \delta \nbr{{\bb_{i_j}}}$. 
\end{lemma}
Let $U_{i_j}'$ be the top $n/2k$ coordinates of the projected vector $\nbr{P_W (\aa_{i_j} - \bar{\aa})}$. If we choose the constants $C_1,C_2$ appropriately, then for every good index $i_j$,  $U_{i_j}$ contains  $0.95$ fraction of the vertices in $V_{n_{i_j}} \cap Z$.



Lemma \ref{lem:testingGood} then implies that when we throw away half of the sets $U_1',...,U_m'$ with the least Blue edge densities, then each of the remaining sets intersects some $V_{n_{i}} \cap Z$ in $0.9$ fraction of the vertices.
\begin{lemma}
\label{lem:testingGood}
There exists a constant $c>0$ such that the following holds. Suppose we are given a set $X \subset Z$ of size $|X| = n/2k$. If for all $i \in {1,...,k}$  $$|X \cap V_{i}| \le 0.9 |X|,$$ then with probability at least $1 - e^{-cn}$  the number of Blue edges in the graph induced by $X$ is at most $an/16k - 0.09(a-b)n/16k$. Conversely, if $$|X \cap V_{i}| \ge 0.95 |X|$$ for some $i \in {1,...,k}$, then with with probability at least $1 - e^{-cn}$ the number of Blue edges in the graph induced by $X$ is at least $an/16k - 0.09(a-b)n/16k.$
\end{lemma}

\proof

Let $e(X)$ denote the number of Blue edges in the graph induced by vertices in $X$. Suppose $|X \cap V_{i}| \le 0.9 |X|$ for all $i \in {1,...,k}$.  Then 
$$\E e(X) \le an/16k - 0.09(a-b)n/8k.$$
To bound the probability that $\E e(X) \ge an/16k - .045(a-b)n/8k$, we can use Chernoff bound. Let $\delta = \frac{0.045(a-b)n/8k}{an/16k - 0.09(a-b)n/8k}.$
$$\mathbb{P}(\E e(X) \ge an/16k - .045(a-b)n/8k) \le \exp \left( -\frac{(.045(a-b)n/8k)^2}{2an/16k+0.045(a-b)n/8k} \right).$$

Similarly, suppose $|X \cap V_{i}| \ge 0.95 |X|$ for some $i \in {1,...,k}$.  Then 
$$\E e(X) \ge an/16k - 0.05(a-b)n/16k.$$
To bound the probability that $\E e(X) \le an/16k - .045(a-b)n/8k$, we can use Chernoff bound. Let $\delta = \frac{0.04(a-b)n/16k}{an/16k - 0.05(a-b)n/16k}.$
$$\mathbb{P}(\E e(X) \le an/16k - .045(a-b)n/8k) \le \exp \left( -\frac{(.04(a-b)n/16k)^2}{2an/16k+0.04(a-b)n/16k} \right).$$

 

\subsection{Proof of lemma \ref{lem:correction_multi}} 
\label{app:multCorrection}

For notional convenience, let $U_i := Z \cap V_i$. We will use the following large deviation result (see page 36 in \cite{boucheron} for e.g.) repeatedly

\begin{lemma} (Chernoff) If $X$ is a sum of $n$ iid indicator random variables with mean at most  $\rho \le 1/2$, then for any $t >0$

$$\max \{ \mathbb{P} (X \ge \E X +t) , \mathbb{P} (X \le \E X -t ) \} \le \exp \left( -\frac{t^2}{2 \var X + t} \right) \le \exp \left( -\frac{t^2}{ 2 n \rho +t  }\right)  . $$ 

\end{lemma} 

  By Theorem {step1}, there is a constant $C$ such that if  $\frac{(a-b)^2 }{a+b} \geq C$  then by running {\bf Spectral Partition} on the Red graph, we obtain,  with probability $1-o(1)$, sets $ U_1',...,U_k' $, where 

$$|U_i' \backslash U_i| \leq 0.1 n/2k.$$

In the rest, we condition on this event. The probability we will talk about in this section is based on the edges that go between vertices in $Z$. 

Now we use the edges that go between vertices in $Z $. Consider $e= (u,v)$. If $u  \in U_i, v \in U_{j}$ with $i \neq j$, then $e$ is a Red edge with probability 

$$\mu: =   b/2n  . $$

Similarly, if  $u , v \in U_i$, then $e$ is a Red edge with probability 

$$\tau : =  a/2n  . $$

For any $u \in  U_1$,  the number of  its  neighbors in $U_j'$ is at most 

$$S_{1i}(u) := \sum_{i=1}^{.9n/2k} \xi_{i}^u   + \sum_{j=1}^{.1 n/2k} \zeta_{ j} ^u $$

Similarly, for any $u \in U_1$, the number of  its  neighbors in $U_1'$ is at  least 

$$S'_{11}(u) := \sum_{i=1}^{.9n/2k } \zeta_{i}^u   + \sum_{j=1}^{.1 n/2k} \xi_{ j} ^u.$$

After the correction sub-routine, if a vertex $u \in U_1$ is mislabeled then one of the following holds

\begin{itemize} 

\item  $S_{1j}' \ge \frac{a+b} {8}  $  for some $j \neq 1$

\item  $S_{11} \le \frac{a+b} {8}  $.

\end{itemize} 
By an application of Chernoff bound, probability that $S_{11} \le \frac{a+b} {8}  $  can be bounded by $\rho_1 = \exp ( - 0.04 \frac{(a-b)^2 }{k(a+b)} ).$  Similarly, for any  fixed $j \neq 1$, $S_{1j}' \ge \frac{a+b} {8} $  is bounded by $\rho_1.$ Therefore, the probability that any of these happens is bounded by $k \rho_1$. Therefore, number of vertices in $U_1$ that will be misclassified after the correction step is at most
$$ M: = \sum_{ k=1}^{n/2k}   \Gamma_k  $$
where $\Gamma_k$ are iid indicator random variables with mean $\rho_1$.

$$\E M \le \frac{n}{2k} k\exp(- 0.04 \frac{(a-b)^2 }{k(a+b)} ). $$

Applying Chernoff's with $t : =\frac{n}{2}  \exp(- 0.04 \frac{(a-b)^2 }{k(a+b)} )$, we conclude that with probability $1-o(1)$
$$M  \le \E M+t  = n \exp( -0.04 \frac{(a-b)^2 }{k(a+b)} ). $$
This implies that with probability $1-o(1)$, number of mislabeled vertices in $U_1$ is 
$$ \le  n \exp( - 0.04 \frac{(a-b)^2 }{k(a+b)} ). $$
 
Set $$\gamma := 2k \exp( - 0.04 \frac{(a-b)^2 }{k(a+b)} ).$$ 
Therefore, by a union bound over all $i$, we have that with probability $1-o(1)$ the output $U_1', U_2',...,U_k'$ after the correction step form a $\gamma$-correct partition, with $\gamma$ satisfying 
$$ \frac{(a-b)^2 }{k(a+b)}=\frac{1}{0.04} \log \frac{2k}{\gamma} = 25 \log \frac{2k}{\gamma} , $$ proving our claim.

\subsection {Proof of lemma \ref{lem:merge_multi}}
\label{app:multMerge}
In this section, we show how we can merge $V_i \cap Y$ with $V_i \cap Z$ based on the Blue edges that go in between vertices in $Y$ and $Z$. We can assume that that we are given a $\gamma$ correct partition $U'_1,..,U'_k$ of $U_1,...,U_k$. Now we label the vertices in $Y$ according to their degrees to $U'_i$ as given in the {\bf Merge} routine.  Let us assume $\gamma \leq 0.1$. In the rest, we condition on this event, and  the event that the maximum Red degree of a vertex is at most $\log^2 n$, which occurs with probability $1-o(1)$.

Now we use the Blue edges. Consider $e= (u,v)$. If $e$ is not a red  edge, and $u  \in V_i \cap Y, v \in V_{j} \cap Z  $, then $e$ is a Blue edge with probability 

$$\mu: =  \frac{ b/2n }{ 1 - \frac{b}{2n }} . $$

Similarly, if $e$ is not a Red edge, and $u \in V_i \cap Z, v \in V_i \cap Z$, then $e$ is a Blue edge with probability 

$$\tau : =  \frac{ a/2n }{ 1-  \frac{a}{2n }} . $$

Thus, for any $u \in Y \cap V_i$,  the number of  Blue neighbors in $U_j'$ is at most 

$$S_j := \sum_{i=1}^{.9n/2k} \xi_{i}^u   + \sum_{j=1}^{.1 n/2k} \zeta_{ j} ^u $$

\noindent  where $\xi_{i}^u $ are iid indicator variables with mean $\mu$ and $\zeta_{j}^u$ are iid indicator variables with mean $\tau$.

Similarly, for any $u \in Y \cap V_i$, the number of  Blue  neighbors in $U_i'$ is at  least 

$$S'_i := \sum_{i=1}^{.9n/2k - d(u) } \zeta_{i}^u   + \sum_{j=1}^{.1 n/2k} \xi_{ j} ^u.$$

After the correction sub-routine, if a vertex $u $ in $ Y \cap V_i$ is misclassified then  one of the following holds

\begin{itemize} 
\item  $S_j  \ge \frac{a+b}{8k} $ 
\item  $S_i' \le \frac{a+b} {8k}  .$
\end{itemize}

Let $\rho$ be the probability that at least one of the above events happens. Then the number of mislabeled vertices in the $Y_2$ is at most 
$$ M: = \sum_{ k=1}^{n/2}   \Gamma_k  $$ where $\Gamma_k$ are iid indicator random variables with mean $\rho$. First we use Chernoff bound to estimate $\rho$.  Consider 

$$\rho_1 := \mathbb{P} \left( S_j \ge \frac{a+b}{8k} \right). $$

By definition, we have

\begin{equation} \label{expectation} 
\begin{split}
\E S(u) &= \frac{0.9n \mu/k + 0.1n \tau/k}{2} \\
&= 0.9 n(\frac{ b/4kn }{ 1 - \frac{b}{2n }} ) +0.1 n(\frac{ a/4kn }{ 1-  \frac{a}{2n }} ) \\
&=  0.9 \frac{b}{4k} + 0.1 \frac{a}{4k} + 0.9 \frac{b}{4k} ( \frac{1}{1- b/2n } -1) + 0.1 \frac{a}{4k} ( \frac{1}{1-a/2n} -1). 
\end{split}
\end{equation}

Set 

$$t:= \frac{a+b}{8k} - \E S_j, $$ we have  

$$t = 0.1 \frac{a-b}{k} - 0.9 \frac{b}{4k} ( \frac{1}{1- b/2n } -1) - 0.1 \frac{a}{4k} ( \frac{1}{1-a/2n} -1) \ge 0.1 \frac{a-b}{k} - 0.9 \frac{b}{4k} \frac{b}{n} - 0.1 \frac{a}{4k} \frac{a}{n} \ge 0.09 \frac{a-b}{k} , $$ for any sufficiently large $n$. 

 Applying Chernoff's bound, we obtain 

\begin{align*}
\rho_1 &\le \exp( - \frac{ (0.09 (a-b) )^2 }{  k(0.9b/2 + .1 a/2) + 0.09 k(a-b) } ) \\
&\leq \exp( - \frac{  0.0324(a-b)^2 }{  k(a+b) } ).
 \end{align*}

 By a similar argument, we get the same bound for $$\rho_2 := \mathbb{P} \left( S'_i \le \frac{a+b}{8k} \right). $$ Therefore, by a union bound, we have that $$\rho \leq  k \exp( - \frac{  0.0324(a-b)^2 }{  k(a+b) } ).$$ Thus, we can conclude that 

$$\E M \le  \frac{n}{2} k\exp(- 0.0324 \frac{(a-b)^2 }{k(a+b)} ). $$

Applying Chernoff's with $t : = \frac{n}{2} k \exp \left(- 0.0324 \frac{(a-b)^2 }{k(a+b)} \right)$, we conclude that with probability $1-o(1)$

$$M  \le \E M+t  = n k\exp \left( -0.0324 \frac{(a-b)^2 }{k(a+b)} \right). $$

This implies that with probability $1-o(1)$, the number of mislabeled vertices in $Y$ is bounded by

$$ nk \exp \left( - 0.0324 \frac{(a-b)^2 }{k(a+b)} \right). $$ 

Set

$$\gamma:  = 2k \exp \left( - 0.0324 \frac{(a-b)^2 }{k(a+b)} \right).$$ We have, with probability $1 - o(1)$, $\gamma$ correct partition of the vertices in $Y$, with $\gamma$ satisfying $$\frac{(a-b)^2 }{k(a+b)} = \frac{1}{0.0324} \log \frac{2k}{\gamma} \leq 31 \log \frac{2k}{\gamma} , $$ proving our claim.

\section{Censor Block Model}
\label{app:CBM}

All we have to do now is to bound $\nbr{E}$. 
Let $\sigma^2 := p  \geq \text{Var}(\zeta_{i,j})$ for all $(i,j)$. $\YY_0$ is obtained by  zeroing out rows and columns of $Y$ of high degree. We then have the following lemma. The proof is essentially the same as corollary~\ref{lem:normE}, so we skip the details.

\begin{lemma}
 $0< \epsilon_0 \leq \epsilon < \frac{1}{2}$. Then there exist constants $C,C_1$ such that if $p  \geq \frac{C}{n}$, then with probability $1-o(1) $, $\nbr{Y_0 - \bar{Y}} \leq C_1 \sigma \sqrt{n}= C_1 \sqrt{np} $.
\end{lemma}

Since the second eigenvalue of $\bar{Y}$ is $p(1 - 2\epsilon)n$, to make the angle between the eigenspace spanned by the two eigenvectors corresponding to the top two eigenvalues small, we need to assume $$\frac{p(1 - 2\epsilon)n }{\sqrt{np} }$$ is sufficiently large. The assumption  $$np \geq \frac{C_2}{(1-2\epsilon)^2}$$ in theorem~\ref{thm:CBM} is precisely this.

\section{Proof of Lemma~\ref{lem:FE_heavy}}
\label{sec:proofFE}
 This proof is essentially same as that in \cite{RSA:RSA20089}. Let us first define the following sets. For $\gamma_k := 2^k$, 
 $$ S_k := \left \{ i: \frac{\gamma_{k-1}}{\sqrt{n}} < x_i \leq   \frac{\gamma_k}{\sqrt{n}} \right \}, s_k := |S_k|, k=\lfloor \log \frac{\sqrt{d}}{n} \rfloor,..,0,1,2,...,\lceil \log \sqrt{n} \rceil$$ and 
 $$ T_k := \left \{ i: \frac{\gamma_{k-1}}{\sqrt{n}} < y_i \leq   \frac{\gamma_k}{\sqrt{n}} \right \}, t_k := |T_k|, k=\lfloor \log \frac{\sqrt{d}}{n} \rfloor,..,0,1,2,...,\lceil \log \sqrt{n} \rceil.$$ Further, we use the notation $\mu_{i,j} := s_i t_j \frac{d}{n}$ and $\lambda_{i,j} := e(S_i,T_j)/ \mu_{i,j}.$ We then have 
 \begin{align*}
 \sum_H x_i \widetilde{\mata}_{i,j} y_j &\leq \sum_{i,j: \gamma_i \gamma_j \geq \sqrt{d}} s_i t_j \frac{d}{n} \lambda_{i,j} \frac{\gamma_i}{\sqrt{n}} \frac{\gamma_j}{\sqrt{n}} \\
 &= \sqrt{d} \sum_{i,j: \gamma_i \gamma_j \geq \sqrt{d}}  s_i \frac{\gamma_i^2}{n} t_j \frac{\gamma_j^2}{n} \frac{\lambda_{i,j} \sqrt{d}}{\gamma_i \gamma_j}\\
 &= \sqrt{d} \sum_{i,j: \gamma_i \gamma_j \geq \sqrt{d}}  \alpha_i \beta_j \sigma_{i,j}.
 \end{align*}

 In the last line, we have used the following notation $\alpha_i := s_i \frac{\gamma_i^2}{n} , \beta_j := t_j \frac{\gamma_j^2}{n}, \sigma_{i,j}:= \frac{\lambda_{i,j} \sqrt{d}}{\gamma_i \gamma_j}.$ In this notation, we can write \ref{enum:descrep2} as follows:
 \begin{align}
 \label{enum:descrep2'}
 \sigma_{i,j} \alpha_i \log \lambda_{i,j} \leq c_3 \frac{\gamma_i}{\gamma_j \sqrt{d}} \left[ 2 \log \gamma_j + \log \frac{1}{\beta_j} \right].
 \end{align}
 
  Now we bound $\sum_{i,j: \gamma_i \gamma_j \geq \sqrt{d}}  \alpha_i \beta_j \sigma_{i,j}$ by a constant. We note that $\sum_i \alpha_i \leq 4$ and $\sum_i \beta_i \leq 4.$ We now consider 6 cases.
 \begin{enumerate}
 \item $\sigma_{i,j} \leq 1:$   
 \begin{align*}
 \sqrt{d} \sum_{i,j: \gamma_i \gamma_j \geq \sqrt{d}}  \alpha_i \beta_j \sigma_{i,j} & \leq \sum_{i,j: \gamma_i \gamma_j \geq \sqrt{d}}  \alpha_i \beta_j \\ 
 &\leq \sqrt{d} (\sum_i \alpha_i)(\sum_i \beta_i)\\
 & \leq 16 \sqrt{d} .
 \end{align*}
 
 \item $\lambda_{ij} \leq c_2:$ Since $\gamma_i \gamma_j \geq \sqrt{d}$ we have in this case $\sigma_{i,j} \leq c_2.$ Therefore,
 \begin{align*}
 \sqrt{d} \sum_{i,j: \gamma_i \gamma_j \geq \sqrt{d}}  \alpha_i \beta_j \sigma_{i,j} & \leq \sum_{i,j: \gamma_i \gamma_j \geq \sqrt{d}}  \alpha_i \beta_j c_2 \\ 
 &\leq c_2 \sqrt{d} (\sum_i \alpha_i)(\sum_i \beta_i)\\
 & \leq 16 c_2 \sqrt{d} .
 \end{align*}
 
 \item $\gamma_i > \sqrt{d} \gamma_j :$ Since the maximum degree is $\leq c_1d$, we have that $\lambda_{i,j} \leq c_1 n/t_j$. Therefore, 
 \begin{align*}
 \sqrt{d} \sum_{i,j: \gamma_i \gamma_j \geq \sqrt{d}}  \alpha_i \beta_j \sigma_{i,j} & =\sqrt{d} \sum_i \left( \alpha_i \sum_{j: \gamma_i \gamma_j \geq \sqrt{d}} \beta_j \frac{\lambda_{i,j} \sqrt{d}}{\gamma_i \gamma_j} \right) \\
 & \leq \sqrt{d}  \sum_i \left( \alpha_i \sum_{j: \gamma_i \gamma_j \geq \sqrt{d}} b_j \frac{\gamma_j^2}{n} \frac{ (c_1 n/b_j) \sqrt{d}}{\gamma_i \gamma_j} \right)\\
 & = \sqrt{d} \sum_i \left( \alpha_i \sum_{j: \gamma_i \gamma_j \geq \sqrt{d}} c_1 \sqrt{d} \frac{\gamma_j}{\gamma_i} \right)\\
 & \leq  \sqrt{d} \sum_i ( \alpha_i c_1  \times 2)\\
 & \leq 2c_1 \sqrt{d} \sum_i \alpha_i \\
 & \leq 8c_1 \sqrt{d}.
 \end{align*}
 
 \item We now assume that we are {\bf not} in cases $1-3$. Therefore, we can assume that \ref{enum:descrep2'} holds. We consider the following sub cases.
 \begin{enumerate}
 \allowdisplaybreaks
 \item $\log \lambda_{i,j} > (1/4) [2 \log \gamma_j  + \log (1/\beta_{j}) ]:$ \ref{enum:descrep2'} implies that $\sigma_{i,j} \alpha_i \leq 4 c_3 (\gamma_i/ \gamma_j \sqrt{d})$. Therefore, 
 \begin{align*}
 \sqrt{d} \sum_{i,j: \gamma_i \gamma_j \geq \sqrt{d}}  \alpha_i \beta_j \sigma_{i,j} & = \sqrt{d} \sum_j \beta_j  \sum_{i: \gamma_i \gamma_j \geq \sqrt{d}}  \alpha_i \sigma_{i,j} \\
 & \leq  \sqrt{d} \sum_j \beta_j   \sum_{i: \gamma_i \gamma_j \geq \sqrt{d}} 4 c_3 \frac{\gamma_i}{\gamma_j \sqrt{d}}\\
 & \leq \sqrt{d}  \sum_j \beta_j  \times 8 c_3\\
 & \leq 32 c_3 \sqrt{d}.
 \end{align*}
 Above we made use of the fact that we are not in case $3$, and that $\sum_{i: \gamma_i \gamma_j \geq \sqrt{d}} 4 c_3 \frac{\gamma_i}{\gamma_j \sqrt{d}}$ is a geometric sum.
 
 \item $2 \log \gamma_j \geq \log (1/\beta_j):$ We can assume we are not in case (a), and hence  $\lambda_{i,j} \leq \gamma_j$. Combined with the fact that we are not in case $1$, we have that $\gamma_i \leq \sqrt{d}$. Since we are not in case $2$, we can assume that $\log \lambda_{i,j} \geq 1$ and hence $\sigma_{i,j} \alpha_i \leq c_3  \frac{\gamma_i}{\gamma_j \sqrt{d}} 4 \log \gamma_j.$ Therefore,
 \begin{align*}
 \sqrt{d} \sum_{i,j: \gamma_i \gamma_j \geq \sqrt{d}}  \alpha_i \beta_j \sigma_{i,j} & = \sqrt{d} \sum_j \left( \beta_j \sum_{i: \gamma_i \gamma_j \geq \sqrt{d}} \alpha_i \frac{\lambda_{i,j} \sqrt{d}}{\gamma_i \gamma_j} \right)\\
 & \leq \sqrt{d} \sum_j \left( \beta_j \sum_{i: \gamma_i \gamma_j \geq \sqrt{d}} 4 c_3 \frac{\gamma_{i}}{ \sqrt{d} \gamma_j} \log \gamma_j \right)\\
 & \leq \sqrt{d}  \left( \sum_j  \beta_j 4 c_3\sum_{i: \gamma_i \gamma_j \geq \sqrt{d}} \frac{\gamma_i}{\sqrt{d}} \right) \\
 & \leq \sqrt{d}  \sum_j  \beta_j 4 c_3 \times 2 \\
 & \leq 32 c_3 \sqrt{d}.
  \end{align*}
 \item $2 \log \gamma_j \leq \log 1/\beta_j:$ Since we are not in $(a)$ we have $\log \lambda_{i,j} \leq \log \frac{1}{\beta_j}$. It follows that 
 $$\sigma_{i,j} = \frac{\lambda_{i,j} \sqrt{d}}{\gamma_i \gamma_j} \leq \frac{1}{\beta_j} \frac{\sqrt{d}}{\gamma_i \gamma_j}.$$ Therefore:


  \begin{align*}
  \allowdisplaybreaks
 \sqrt{d} \sum_{i,j: \gamma_i \gamma_j \geq \sqrt{d}}  \alpha_i \beta_j \sigma_{i,j} & = \sqrt{d} \sum_i \left( \alpha_i \sum_{j: \gamma_i \gamma_j \geq \sqrt{d}} \beta_j \sigma_{i,j} \right)  \\
 &= \sqrt{d} \sum_i \left( \alpha_i \sum_{j: \gamma_i \gamma_j \geq \sqrt{d}} \beta_j \frac{\lambda_{i,j} \sqrt{d}}{\gamma_i \gamma_j} \right)  \displaybreak \\
 & \leq  \sqrt{d} \sum_i \left( \alpha_i \sum_{j: \gamma_i \gamma_j \geq \sqrt{d}} \frac{ \sqrt{d}}{\gamma_i \gamma_j} \right) \\
  & \leq  \sqrt{d} \sum_i \left( \alpha_i \times 2\right)\\
  & \leq  8 \sqrt{d}.\\
\end{align*}
 
 \end{enumerate}
 \end{enumerate}
 
 \QED

\end{document}